\documentclass[AMA,Times1COL]{WileyNJDv5} 
\usepackage{caption}
\usepackage{enumitem} 
\usepackage{subcaption}
\articletype{ORIGINAL ARTICLE}%
\received{Date Month Year}
\revised{Date Month Year}
\accepted{Date Month Year}
\journal{Journal}
\volume{00}
\copyyear{2026}
\startpage{1}
\begin{document}
\title{Bayesian inference of risk differences for multi-group bilateral data}

\author[]{Jinxiu Wen | Zhiming Li}

\authormark{Wen \textsc{et al.}}
\titlemark{Bayesian inference of risk differences for multi-group bilateral data}

\address[]{\orgdiv{College of Mathematics and System Science}, \orgname{Xinjiang University}, \orgaddress{\state{Urumqi 830017}, \country{China}}}



\corres{\email{zmli@xju.edu.cn}}

\presentaddress{The work was supported by the National Natural Science Foundation of China (12561047), the Xinjiang Talent Development Fund (XJRC-2025 -KJ-PY-KJLJ-108), and the 2025 Central Guidance for Local Science and Technology Development Fund (ZYYD2025ZY20).}

\abstract[Abstract]{Bilateral data from paired body parts are common and correlated in clinical studies. Classical frequentist methods perform poorly in small or sparse datasets. This paper develops a Bayesian framework for bilateral data with multiple groups under Dallal's model. We derive three objective priors (uniform, Jeffreys', and Bernardo's reference priors) and propose a range-based posterior testing procedure, combined with a decision rule, to test the homogeneity of risk differences, with the equivalence margin calibrated. Monte Carlo simulations evaluate empirical Type I error rates, powers, and interval estimation properties. Results show that the Bayesian methods achieve accurate coverage probabilities, narrower confidence intervals, and better Type I error control than the frequentist Wald approach, especially in small-sample and sparse-data settings. We illustrate the methodology with two real datasets. The proposed framework provides a robust and flexible tool for multi-group bilateral data analysis.}



\maketitle


\section{Introduction}\label{Sec1}
Bilateral data arise in medicine when patients receive treatment on paired body parts within the same individual, such as eyes, ears, hands, or knees. The investigator records paired Bernoulli outcomes that are grouped into three categories: both parts cured, exactly one cured, or neither cured (Rosner, 1988)\cite{rosner1988significance}. Such data are common in ophthalmologic, orthopedic, and otolaryngologic studies, as well as in twin studies. The goal of such clinical trials is to quantify the benefit of treatment over control. The dependency between paired observations cannot be ignored, as ignoring it leads to biased inferences (Rosner, 1982\cite{rosner1982statistical}; Morris, 1993\cite{morris1993bilateral}; Tang et al., 2008\cite{tang2008testing}). Dallal's model (Dallal, 1988\cite{dallal1988paired}) characterizes this dependency through a single correlation parameter \(\gamma\), which has a clear biological interpretation and is among the few bilateral models that admit closed-form objective priors (M'lan and Chen, 2015\cite{mlan2015objective}). We therefore focus on Dallal's model. While Dallal's model specifies the correlation structure of bilateral data through a single dependence parameter \(\gamma\) (Dallal, 1988)\cite{dallal1988paired}, our primary inferential goal is the homogeneity of risk differences across groups. This is consistent with the objective Bayesian framework of M'lan and Chen (2015)\cite{mlan2015objective}, who derived closed-form objective priors for the risk difference under Dallal's model, and with recent frequentist work on risk difference homogeneity testing (Sun et al., 2022\cite{Sun2022risk}; Zhao et al., 2023\cite{Zhao2023homogeneity}).

A substantial body of literature has addressed homogeneity testing for multi-group bilateral data, including correlation homogeneity and risk-difference homogeneity. Li et al. (2020)\cite{li2020statistical} proposed asymptotic test statistics and exact procedures for testing the equality of correlations among multiple bilateral data, and found that the score test is the most robust in large samples while exact methods are needed for small samples. More recently, Liang et al. (2024)\cite{liang2024homogeneity} conducted homogeneity tests and confidence intervals of risk differences for stratified correlated data, and Sun et al. (2025)\cite{Sun2025interval} constructed confidence intervals for the common risk difference under stratified unilateral and bilateral data. Hua and Ma (2024)\cite{Hua2024common} proposed large-sample tests and confidence interval methods for the common odds ratio in stratified designs. Liu et al. (2025)\cite{Liu2025Testing} proposed likelihood ratio, score, and Wald-type statistics for testing the equality of response rate functions, which encompass homogeneity tests of risk difference, relative risk ratio, and odds ratio as special cases. Zhou and Ma (2025)\cite{zhou2025testing} developed three likelihood-based test statistics for the risk difference under Donner's constant correlation model. Zhao et al. (2023)\cite{Zhao2023homogeneity} extended the homogeneity test of response rate functions to the general case of \(g\) groups under Dallal's model, covering risk difference, relative risk ratio, and odds ratio. These works provide a comprehensive frequentist toolkit for multi-group bilateral data. However, they rely on asymptotic approximations, which may perform poorly with small sample sizes or sparse data. In such settings, empirical type I error rates can be inflated, and confidence intervals may have poor coverage.

Bayesian methods offer a reliable alternative. They do not rely on the normal approximation, which is a distinct advantage for multi-group bilateral data where the dimension of the parameter space grows with the number of groups. They also naturally handle sparse data and boundary estimates. In the context of Dallal's model, M'lan and Chen (2015)\cite{mlan2015objective} derived Jeffreys' prior and Bernardo's reference prior for bilateral data and studied objective Bayesian inference for the risk difference, risk ratio, and odds ratio. However, their framework is limited to two groups and does not address hypothesis testing. More recently, Wen and Li (2026)\cite{wen2026bayesian} developed a Bayesian framework for multinomial logistic regression based on Dallal's model for stratified unilateral and bilateral paired data, using the No-U-Turn Sampler (NUTS) and Hamiltonian Monte Carlo (HMC) algorithms for posterior sampling and Bayes factors via the Savage--Dickey density ratio. Their results showed that the Bayesian approach outperforms traditional frequentist methods, especially at small sample sizes. However, their method targets regression coefficients under a logistic model and does not address the homogeneity of risk differences across multiple groups. To the best of our knowledge, no Bayesian hypothesis testing method has been developed for multi-group bilateral data under Dallal's model.

In this paper, we propose a Bayesian framework for multi-group bilateral data under Dallal's model and develop a posterior range test for the homogeneity of risk differences. We derive three priors, namely the uniform, Jeffreys' and Bernardo's reference priors. For hypothesis testing, we adopt the HDI+ROPE decision rule, which combines the highest density interval (HDI) with a region of practical equivalence (ROPE), and calibrate the equivalence margin for each combination of group number, sample size, and prior. We also compute Bayes factors via the Savage--Dickey density ratio to provide complementary evidence for the homogeneity test. Simulation studies are performed to evaluate interval estimation properties, empirical type I error rates, and empirical powers. Two real datasets illustrate the methodology: a two-group scleroderma trial that validates the reduction to the original Dallal's model, and a four-group retinitis pigmentosa study that demonstrates applicability to unbalanced multi-group designs.

The remainder of this paper is organized as follows. Section \ref{Sec2} presents a multi-group Dallal's model. Section \ref{Sec3} derives the objective priors and posterior sampling. Section \ref{Sec4} proposes the posterior range test and calibration. Section \ref{Sec5} reports simulation results. Section \ref{Sec6} illustrates with real data. Section \ref{Sec7} concludes.

\section{Dallal's model with multiple groups}\label{Sec2}
Suppose all patients are divided into \(g(\geq 2)\) groups. Denote \(M\) as the patients that contribute a pair of organs. Let \(m_{i}=\sum_{l=0}^2 m_{lg}\) (\(l=0, 1, 2\)) be the total number of bilateral patients respectively in the \(i\)th (\(i=1, 2,\dots, g\)) group, where \(m_{li}\) represents the number of patients having \(l\) response(s). Obviously, \(M=\sum_{i=1}^2 m_{i}\). The detailed data structure is presented in Table \ref{tab:data_structure0}.
\begin{table*}[!htbp]
	\centering
	\caption{Data structure of bilateral paired data from \(g\) groups.}
	\label{tab:data_structure0}
	\begin{tabular*}{\textwidth}{@{\extracolsep\fill}llccccc@{\extracolsep\fill}}
		\toprule
		\multirow{2}{*}{Data Type} & \multirow{2}{*}{Number of Responses (\(l\))} & \multicolumn{4}{c}{Group ($i=1,2,\cdots,g$)} & \multirow{2}{*}{Total} \\
		\cmidrule{3-6}  
		& &1 &2 &$\cdots$ & \(g\)  \\
		\midrule
		\multirow{4}{*}{Bilateral} 
		& 0   & $m_{01}$ & $m_{02}$ &$\cdots$ &$m_{0g}$ & $M_{0}$ \\
		& 1   & $m_{11}$ & $m_{12}$ &$\cdots$ &$m_{1g}$ &  $M_{1}$ \\
		& 2   & $m_{21}$ & $m_{22}$ &$\cdots$ &$m_{2g}$ &  $M_{2}$ \\
		& Total& $m_{1}$ & $m_{2}$ &$\cdots$ &$m_{g}$ &  $M$ \\
		\bottomrule
	\end{tabular*}
\end{table*}

According to Dallal's model, we denote \(Z_{ij}=1\) as the event that the \(k\)th (\(k=1, 2\)) organ studied of the \( j \)th (\(j=1, \cdots, m_{i}\)) patient from the \(g\)th group has one response and \(Z_{ij}=0\) otherwise. Also, the probabilities for none, one, and two responses are \(P_{0i}\), \(P_{1i}\), and \(P_{2i}\). Then, the probability of having one response is described by \(\pi_{i}\) for our data. Let \(1-\gamma_{i}\) denote the conditional probability of occurrence at one side when a response occurs at the other side. Thus, Dallal's model can be characterized by
$
		Pr(Z_{ijk}=1)=\pi_{i},  
		Pr(Z_{ijk}=1\mid Z_{ij(3-k)} =1)=1-\gamma_{i},
$
where \(\pi_{i}, \gamma_{i}\in [0, 1]\). Then, some probabilities are straightforwardly given by
\begin{eqnarray*}
	\begin{aligned}
		P_{0i}=1-(1+\gamma_{i})\pi_{i},\quad P_{1i}=2\gamma_{i}\pi_{i},\quad P_{2i}=(1-\gamma_{i})\pi_{i}.
	\end{aligned}
\end{eqnarray*}
We assume that the conditional probability of the same characteristic occurring at the other site given the occurrence of a certain characteristic at one site is equal in each treatment group in this paper, i.e., \(\gamma_{i}=\gamma\). Importantly, the probabilities satisfy the normalization condition so that the parameter space is constrained; the parameter space is
\begin{eqnarray*}
	\begin{aligned}
		\Omega = \big\{ (\gamma, \pi_{1}, \cdots, \pi_{g}) \mid 0 \leq \gamma \leq 1 \text{ and } 0 \leq \pi_{i} \leq \frac{1}{1+\gamma},  i=1,\cdots, g\big\}.
	\end{aligned}
\end{eqnarray*}
The observations \(\boldsymbol{m}_{i}=({m}_{0i}, {m}_{1i}, {m}_{2i})^T\) (\(i=1,2,\cdots,g\)) follows a trinomial distribution of the \(i\)th group. Let \(\boldsymbol{D}=(\boldsymbol{m}_1, \cdots, \boldsymbol{m}_g)\) represent all data observed. Hence, the likelihood  function can be expressed as
\begin{eqnarray}\label{likehood_h0}
	\begin{aligned}
		L(\gamma, \boldsymbol{\pi}| \boldsymbol{D})&=\prod_{i=1}^g \frac{m_{i}!}{m_{0i}! \, m_{1i}! \, m_{2i}!} P_{0i}^{m_{0i}} P_{1i}^{m_{1i}} P_{2i}^{m_{2i}} \\ &\propto \gamma^{M_{1}}(1-\gamma)^{M_{2}} \prod_{i=1}^{g}\pi_{i}^{m_{1i}+m_{2i}} [1-(1+\gamma)\pi_{i}]^{m_{0i}},\quad (\gamma, \boldsymbol{\pi}) \in \Omega,
	\end{aligned}
\end{eqnarray}
where \(\boldsymbol{\pi}=(\pi_{1}, \cdots, \pi_{g})\), \(M_{1}=\sum_{i=1}^g m_{1i}\) and \(M_{2}=\sum_{i=1}^g m_{2i}\).

\section{Bayesian Analysis}\label{Sec3}
To conduct Bayesian inference, the main challenge is specifying priors for unknown parameters. Using a proper prior is crucial for deriving the posterior distribution. Traditionally, most Bayesians have turned to conjugate priors due to their computational convenience. However, in multi-parameter problems or when little prior information is available, conjugate priors may become overly subjective and difficult to justify. In this paper, we adopt non-informative or objective priors that convey minimal prior information and let the data dominate the posterior inference. Specifically, we consider three objective priors under the multi-group Dallal's model: the uniform prior, Jeffreys' prior, and Bernardo's reference prior. The uniform prior over the constrained parameter space is the simplest choice. Jeffreys' prior derived from the Fisher information matrix is invariant under one-to-one reparameterization. The reference prior maximizes the information divergence between the prior and the posterior, minimizing the prior's influence on our inference.

In the following subsections, we present the derivation of these three priors under the parameterization $(\gamma, U_1, \dots, U_g)$, where $U_i = (1+\gamma)\pi_i$ for $i=1,\dots,g$. This parameterization can greatly simplify the derivation of Jeffreys' prior and the reference prior. For each prior, we provide the closed-form expression of the joint prior density as well as the corresponding posterior distribution. We discuss our posterior sampling in Section~\ref{Sec3.4}.

\subsection{The Uniform Prior Distribution}\label{Sec3.1}
The uniform prior  under Dallal's model for \((\gamma,\boldsymbol{\pi})\) is 
$
\pi_U(\gamma,\boldsymbol{\pi}) = \frac{g-1}{1 - 2^{1-g}}$ for $(\gamma, \boldsymbol{\pi}) \in \Omega.
$
The prior is proper. Combined with (\ref{likehood_h0}), the posterior distribution is
\begin{eqnarray}\label{piU}
	\begin{aligned}
		\pi_U(\gamma,\boldsymbol{\pi} \mid \boldsymbol{D}) &\propto \pi_U(\gamma,\boldsymbol{\pi}) \cdot L(\gamma, \boldsymbol{\pi} \mid \boldsymbol{D}) \\
		&\propto \gamma^{M_1}(1-\gamma)^{M_2} \prod_{i=1}^g \pi_i^{m_{1i}+m_{2i}} [1-(1+\gamma)\pi_i]^{m_{0i}},\quad (\gamma, \boldsymbol{\pi}) \in \Omega.
	\end{aligned}
\end{eqnarray}
Based on the (\ref{piU}), the marginal posterior
distribution of the parameter \(\gamma\) is
\begin{eqnarray*}
	\begin{aligned}
		\pi_U(\gamma \mid \boldsymbol{D}) =\frac{2^{M_{1}+g-1}}{\text{B}(M_1+1,M_2+1)} \frac{\gamma^{M_1}(1-\gamma)^{M_2}}{(1+\gamma)^{M_1+M_2+g}}
	\end{aligned}, \quad 0<\gamma<1,
\end{eqnarray*}
and the conditional posterior distribution for \(\pi_i\) is
\begin{eqnarray*}
	\begin{aligned}
		\pi_U(\pi_i \mid \gamma,\boldsymbol{D})&=\frac{\pi_U(\gamma,\boldsymbol{\pi} \mid \boldsymbol{D})}{\pi_U(\gamma \mid \boldsymbol{D})} =\frac{(1+\gamma)^{m_{1i}+m_{2i}+1}}{\text{B}(m_{1i}+m_{2i}+1,m_{0i}+1)} \pi_{i}^{m_{1i}+m_{2i}}[1-(1+\gamma)\pi_{i}]^{m_{0i}},\quad 0<\pi_{i}<\frac{1}{1+\gamma},
	\end{aligned}
\end{eqnarray*}
where \(\text{B}(\cdot|\cdot)\) denotes the Beta function. Let \(V_i=(1+\gamma)\pi_i\), then \(V_i \big| \gamma \sim \text{Be}(m_{1i}+m_{2i}+1, m_{0i}+1)\), \(i=1,\cdots,g\). When \(g=2\), \(\frac{1-\gamma}{1+\gamma} \sim \text{Be}(M_{1}+1,\ M_{2}+1)\), where the notation \(\text{Be}(\cdot|\cdot)\) represents the standard Beta distribution. The uniform prior is simple, but it lacks reparameterization invariance. 

\subsection{Jeffreys' Prior Distribution}\label{Sec3.2}
Jeffreys' prior is invariant under reparameterization. Let \(u_i = (1+\gamma)\,\pi_i\), then \(\pi_i=u_i \big/(1+\gamma)\). Under this new parametrization, \(\boldsymbol{u}=(u_1,\cdots,u_g)\) (\(i=1,\cdots,g\)). The constrained parameter space becomes the unconstrained unit hypercube \((0,1)^{g+1}\) for \((\gamma, \boldsymbol{u})\). The likelihood function is given by
\begin{eqnarray}\label{likehood_h0.1}
	\begin{aligned}
		L(\gamma, \boldsymbol{u} \mid \boldsymbol{D}) \propto \frac{\gamma^{M_{1}}(1-\gamma)^{M_{2}}}{(1+\gamma)^{M_{1}+M_{2}}} \prod_{i=1}^g u_{i}^{m_{1i}+m_{2i}} (1-u_{i})^{m_{0i}},\quad 0<\gamma, u_i <1.
	\end{aligned}
\end{eqnarray}
From (\ref{likehood_h0.1}), we can see that the parameters are orthogonal in the sense of Cox and Reid (1987)\cite{cox1987parameter}, and this parametrization splits the likelihood function of \(g\) groups into two unrelated pieces. This makes it very easy to derive the Jeffreys' prior as follows:
\begin{eqnarray}\label{likehood_h0.2}
	\begin{aligned}
		\ell  = \ell(\gamma)+\sum_{i=1}^{g}\ell(u_i) 
		 =M_{1}\ln\gamma+M_{2}\ln(1-\gamma)-(M_{1}+M_{2})\ln(1+\gamma)+
		\sum_{i=1}^{g}\Big[(m_{1i}+m_{2i})\ln u_i+m_{0i}\ln (1-u_i)\Big].
	\end{aligned}
\end{eqnarray}
Through (\ref{likehood_h0.2}), the Jeffreys' prior is proportional to the square root of the absolute value of the determinant of the Fisher information matrix (Jeffreys, 1946\cite{jeffreys1946invariant}). Thus, the Jeffreys' prior of \((\gamma,\boldsymbol{u})\) is
\begin{eqnarray}\label{Prior_J}
	\begin{aligned}
		\pi_J(\gamma,\boldsymbol{u})
		 \propto
		\sqrt{I_{\gamma}}\prod_{i=1}^g\sqrt{I_{u_i}}   
		 \propto \sqrt{\frac{ \sum_{i=1}^{g}r_{i} u_i}{\gamma(1-\gamma)(1+\gamma)^2 \prod_{i=1}^g u_i(1-u_i)}}, \quad 0<\gamma, u_i<1,
	\end{aligned}
\end{eqnarray}
where 
\begin{eqnarray*}\label{Fisher information}
	\begin{aligned}
		I_{\gamma} = \frac{2m_1 \sum_{i=1}^g r_i u_i}{\gamma(1-\gamma)(1+\gamma)^2}, \quad I_{u_i} = \frac{m_{1}r_i}{u_i(1-u_i)},\quad i=1,2,\dots,g,
	\end{aligned}
\end{eqnarray*}
with \(r_i=\frac{m_i}{m_1}\) (\(i=1, \cdots,g\)) being the ratio of the sample sizes in two groups, then \(m_{i}=m_{1}r_{i}\) and \(r_1=1\). The detailed derivation is given in Appendix \ref{app:A}. Under Jeffreys' prior, \(\gamma\) is independent of \(u_i\). Hence, the marginal prior distribution of \(\gamma\) is given by
$\pi_J(\gamma) \propto \frac{1}{\sqrt{\gamma(1-\gamma)}\,(1+\gamma)}.
$
However, the Jeffreys' prior indirectly depends on the sample size via the ratio \(r_i\). The corresponding posterior distribution is
\begin{eqnarray}\label{Posterior_J_U}
	\begin{aligned}
		\pi_J(\gamma,\boldsymbol u\mid \boldsymbol D) \propto
		\frac{\gamma^{M_{1}-1/2}(1-\gamma)^{M_{2}-1/2}}{(1+\gamma)^{M_{1}+M_{2}+1}} \sqrt{\sum_{i=1}^g r_i u_i}\prod_{i=1}^g u_i^{m_{1i}+m_{2i}-1/2}(1-u_i)^{m_{0i}-1/2}.
	\end{aligned}
\end{eqnarray}
Combining (\ref{Prior_J}) and (\ref{Posterior_J_U}), the marginal posterior distribution of the parameter \(\gamma\) is
\begin{eqnarray*}
	\begin{aligned}
		\pi_J(\gamma \mid \boldsymbol{D})
		\propto
		\frac{\gamma^{M_{1}-1/2}(1-\gamma)^{M_{2}-1/2}}
		{(1+\gamma)^{M_{1}+M_{2}+1}},
		\quad \gamma\in(0,1),
	\end{aligned}
\end{eqnarray*}
and the marginal posterior distribution for \(\boldsymbol{u}\) is
\begin{eqnarray*}
	\begin{aligned}
		\pi_J(\boldsymbol{u} \mid \boldsymbol{D})
		\propto
		\sqrt{\sum_{i=1}^g r_i u_i}
		\prod_{i=1}^g u_i^{m_{1i}+m_{2i}-1/2}(1-u_i)^{m_{0i}-1/2},
		\quad u_i \in(0,1).
	\end{aligned}
\end{eqnarray*}
In the original space, we can obtain the Jeffreys' prior as
\begin{eqnarray}\label{Prior_J_orignal}
	\begin{aligned}
		\pi_J(\gamma,\boldsymbol{\pi})
		\propto \sqrt{\frac{\sum_{i=1}^{g}r_{i} \pi_i}{\gamma(1-\gamma)(1+\gamma)^{1-g} \prod_{i=1}^g \pi_i\big[1-(1+\gamma)\pi_i\big]}}, \quad (\gamma, \boldsymbol{\pi}) \in \Omega,
	\end{aligned}
\end{eqnarray}
and the posterior distribution under the Jeffreys' prior is
\begin{eqnarray}\label{Posterior_J_orignal}
	\begin{aligned}
		\pi_J(\gamma,\boldsymbol{\pi}\mid \boldsymbol{D})
		\propto
		\sqrt{\sum_{i=1}^g r_i \pi_i}\gamma^{M_1-1/2}(1-\gamma)^{M_2-1/2}(1+\gamma)^{\frac{g-1}{2}}
		\prod_{i=1}^g \pi_i^{m_{1i}+m_{2i}-1/2}\big[1-(1+\gamma)\pi_i\big]^{m_{0i}-1/2},\quad (\gamma, \boldsymbol{\pi}) \in \Omega.
	\end{aligned}
\end{eqnarray}

\subsection{Reference prior Distribution}\label{Sec3.3}
Although Jeffreys' prior enjoys parameterization invariance, it often encounters serious difficulties in multiparameter problems (Datta and Ghosh, 1996\cite{Datta1996invariance}). The prior may be difficult to derive and complex to interpret. Under Dallal's model, Jeffreys' prior depends on the ratio of sample sizes \(r_{i}=m_{i}/m_{1}\), which makes it design-specific and lacks generality across different designs. When nuisance parameters are present, Jeffreys' prior can become intractable. Therefore, we further consider Bernardo's reference prior (Bernardo, 1979\cite{bernardo1979reference}; Berger and Bernardo, 1989\cite{berger1989estimating}), which provides an alternative objective prior that is invariant under reparameterization and is constructed by ordering the parameters according to their inferential importance (Berger and Bernardo, 1992\cite{berger1992ordered}). In our model, the reference prior not only does not depend on sample sizes, but also renders \(\gamma\) independent of all \(u_i\). We first derive the reference prior for the parameters \(\gamma\) and \(u_i\), and then obtain the induced reference prior for \(\gamma\) and \(\pi_i\). 

Let \(\pi_R(\gamma\mid\boldsymbol{u})\) be conditional reference prior given \(\boldsymbol{u}\), we have
\begin{eqnarray*}\label{reference_prior}
	\begin{aligned}
		\pi_R(\boldsymbol{u},\gamma) &= \pi_R(\gamma\mid\boldsymbol{u})\pi_R(\boldsymbol{u})  
		 = \frac{\arrowvert I_{\gamma}\arrowvert ^{1/2} }{\int \arrowvert I_\gamma \arrowvert^{1/2} d\gamma} \frac{\exp\left\{ \frac{1}{2} \int \log\left(\prod_{i=1}^g I_{u_i} \right) d\gamma \right\}}{\int \exp\left\{ \frac{1}{2} \int \int \cdots \int \log\left( \prod_{i=1}^g I_{u_i} \right) d\gamma du_1 \cdots du_g \right\}d\gamma}\\
		&\propto \frac{\gamma^{-1/2}(1-\gamma)^{-1/2}}{1+\gamma} \prod_{i=1}^g u_i^{-1/2}(1-u_i)^{-1/2},
	\end{aligned}
\end{eqnarray*}
where \(\exp\left( \frac{1}{2} \int \log\left( \prod I_{u_i} \right) d\gamma \right) \propto \prod_{i=1}^g \sqrt{I_{u_i}}\), \(\exp\left\{ \frac{1}{2} \int \cdots \int \log\left( \prod_{i=1}^g I_{u_i} \right) d\gamma du_1 \cdots du_g \right\}=C\) (C is a constant). Bernardo's reference prior corresponding to the groups ordering \(\{\boldsymbol{u}\}\) and then or ordering \(\{\gamma\}\) and then \(\{\boldsymbol{u}\}\) is
\begin{eqnarray}\label{Prior_R_U}
	\begin{aligned}
		\pi_R(\gamma,\boldsymbol{u}) \propto \frac{\gamma^{-1/2}(1-\gamma)^{-1/2}}{1+\gamma} \prod_{i=1}^g u_i^{-1/2}(1-u_i)^{-1/2},\quad 0<\gamma, u_i<1.
	\end{aligned}
\end{eqnarray}
Based on (\ref{Prior_R_U}), Bernardo's reference prior in the original space is
\begin{eqnarray}\label{Prior_R_pi}
	\begin{aligned}
		\pi_R(\gamma,\boldsymbol{\pi}) \propto \frac{\gamma^{-1/2}(1-\gamma)^{-1/2}}{(1+\gamma)^{(2-g)/2}} \prod_{i=1}^g \pi_i^{-1/2}[1-(1+\gamma)\pi_i]^{-1/2},\quad (\gamma, \boldsymbol{\pi}) \in \Omega.
	\end{aligned}
\end{eqnarray}
The posterior distribution resulting from the use of Bernardo's reference prior is
\begin{eqnarray}\label{Posterior_R_U}
	\begin{aligned}
		\pi_R(\gamma,\boldsymbol{u}\mid D)
		\propto
		\frac{\gamma^{M_{1}-1/2}(1-\gamma)^{M_{2}-1/2}}{(1+\gamma)^{M_{1}+M_{2}+1}}
		\prod_{i=1}^g u_i^{m_{1i}+m_{2i}-1/2}(1-u_i)^{m_{0i}-1/2},
	\end{aligned}
\end{eqnarray}
  the marginal posterior
distribution of the parameter \(\gamma\) is
\begin{eqnarray*}\label{mar_gamma}
	\begin{aligned}
		\pi_R(\gamma \mid \boldsymbol{D})
		\propto
		\frac{\gamma^{M_{1}-1/2}(1-\gamma)^{M_{2}-1/2}}
		{(1+\gamma)^{M_{1}+M_{2}+1}},
		\quad \gamma\in(0,1),
	\end{aligned}
\end{eqnarray*}
and the marginal posterior distribution for \(u_i\) is
\begin{eqnarray*}
	\begin{aligned}
		\pi_R(\boldsymbol{u} \mid \boldsymbol{D}) \propto \prod_{i=1}^g u_i^{m_{1i}+m_{2i}-1/2}(1-u_i)^{m_{0i}-1/2},\quad u_i \in(0,1).
	\end{aligned}
\end{eqnarray*}
Hence, \(u_i | \gamma,m_{i} \sim \text{Be}(m_{1i}+m_{2i}+1/2,m_{0i}+1/2)\).
The posterior distribution obtained under Bernardo's reference prior is
\begin{eqnarray}\label{Posterior_R_orginal}
	\begin{aligned}
		\pi_R(\gamma,\boldsymbol{\pi}\mid D)
		\propto
		\frac{\gamma^{M_{1}-1/2}(1-\gamma)^{M_{2}-1/2}}{(1+\gamma)^{(2-g)/2}}
		\prod_{i=1}^g
		\pi_i^{m_{1i}+m_{2i}-1/2}\big[1-(1+\gamma)\pi_i\big]^{m_{0i}-1/2},\quad (\gamma, \boldsymbol{\pi}) \in \Omega.
	\end{aligned}
\end{eqnarray}

\subsection{Sampling from the  Posterior Distribution}\label{Sec3.4}
When analytical solutions are difficult to derive, the Markov chain Monte Carlo (MCMC) method is often employed\cite{murphy2023}. We first discuss how to generate \(\gamma\) from the following distribution
\begin{eqnarray}\label{mar_posterior}
	\begin{aligned}
		f(\gamma)\propto \frac{\gamma^{\mu-1}(1-\gamma)^{\nu-1}}{(1+\gamma)^{\mu+\nu}},\quad \gamma \in(0,1),
	\end{aligned}
\end{eqnarray}
which includes the marginal posterior distributions. We have \(\pi_J(\gamma \mid \boldsymbol{D})=\pi_R(\gamma \mid \boldsymbol{D})\) with \(\mu=M_{1}+1/2\) and \(\nu=M_{2}+1/2\). For the uniform prior, when \(g=2\), \(\pi_U(\gamma \mid \boldsymbol{D})\) is a special case with \(\mu=M_{1}+1\) and \(\nu=M_{2}+1\). However, the marginal posterior of \(\gamma\) when \(g\geq3\) does not satisfy the distribution (\ref{mar_posterior}).
The approach proposed to simulate \(T\) observations \(\gamma_t\) is as follows:
\begin{enumerate}[]
	\item[(i)] Let \(\gamma =  \frac{1-\varphi}{1+\varphi}\), where  \(\varphi \sim \text{Be}(\nu,\mu)\).
	\item[(ii)] Generate \(\varphi_t \sim \text{Be}(\nu,\mu)\), then compute \(\gamma_t =  \frac{1-\varphi_t}{1+\varphi_t}\), \(t=1,\cdots,T\),
\end{enumerate} 
where \(\mu=M_{1}+1/2 ,\nu=M_{2}+1/2\) under the Jeffreys' prior and reference prior. 

Under the Jeffreys' prior, the joint marginal posterior distribution for \((u_1,\cdots,u_g)\) is 
\begin{eqnarray*}
	\begin{aligned}
		\pi_J(u_1,\cdots,u_g \mid \boldsymbol{D})
		\propto
		\sqrt{\sum_{i=1}^g r_i u_i}
		\prod_{i=1}^g u_i^{m_{1i}+m_{2i}-1/2}(1-u_i)^{m_{0i}-1/2},
		\quad u_i \in(0,1).
	\end{aligned}
\end{eqnarray*}
The groups \((u_1,\cdots,u_g)\) are not independent, but are independent of \(\gamma\). We now discuss how to simulate \((u_1,\cdots,u_g)\)
\begin{enumerate}[]
	\item[(i)]  Independently generate \(u_{it}\sim\text{Be}(m_{1i}+m_{2i}+1/2,m_{0i}+1/2)\), \(i=1,\cdots,g\); \(t=1,\cdots,T\).
	\item[(ii)]  Compute the weight \(w_t=\sqrt{\sum_{i=1}^g r_{i} u_{it}}, r_i=\frac{m_{i}}{m_{1}}\).
	\item[(iii)]  Use the acceptance-rejection sampling: Draw \(\xi_{t}\sim U(0,1)\)and accept \(\boldsymbol{u}\) if \(\xi_{t}<w_{t}\big/\sqrt{\sum_{i=1}^{g}r_{i}}\).
	\item[(iv)]  Or use the importance sampling method, where all the \(\boldsymbol{u}\) are accepted and use the weights \(w_t\) to correct for the bias in the computation of posterior mean, quantiles, and highest posterior density (HPD) intervals.
\end{enumerate} 

Under the reference prior, we simulate independent \(T\) observations \((u_1,\cdots,u_g)\) with \(u_i \sim   (m_{1i}+m_{2i}+1/2,m_{0i}+1/2)\). Having simulated \((\gamma_t, \boldsymbol{u}_t)\), we could compute \(\pi_{it} = u_{it}\big/(1+\gamma_t), i=1,\dots,g\) and risk difference \(\delta_{it}=(u_{it}-u_{1t})\big/(1+\gamma_t)\), as well as the the risk ratio \(R_{it}=u_{it}\big/u_{1t}\) and the odds ratio \(\psi_{it}=u_{it}(1+\gamma_{t}-u_{1t})\big/ [u_{1t}(1+\gamma_{t}-u_{it})]\). We use these simulated values to compute posterior probabilities and Bayesian credible intervals, such as equal-tailed intervals and HPD intervals. Bayesian HPD intervals are the shortest intervals containing the parameter of interest with the desired posterior coverage probability. They are more desirable than the commonly used equal-tailed intervals when the posterior distribution is highly asymmetric or skewed, but are more difficult to compute. Chen and Shao (1999)\cite{chen1999monte} develop the Monte Carlo method to compute the HPD intervals under MCMC sampling and importance sampling. Joshi et al. (2023)\cite{joshi2023bayesian} proposed Bayesian simultaneous credible intervals for effect measures from multiple markers, including both equal-tailed and HPD intervals based on the joint posterior distribution.

\section{Bayesian Hypothesis Testing}\label{Sec4}
Let \(\pi_1\) denote the response rate of the control group, and \(\pi_2,\dots,\pi_g\) denote the response rates of the treatment groups. Define the risk difference for the $i$-th treatment group relative to the control group as \(\delta_i = \pi_i - \pi_1 (i = 2,3,\dots,g)\). In this paper, the primary hypothesis of interest is whether the risk differences are equal across all groups in paired data. The
problem can be solved by analyzing the homogeneity test
\begin{equation}
	H_0: \delta_2 = \delta_3 = \dots = \delta_g \triangleq \delta, \quad H_1: \exists\, r \neq s,\; \delta_r \neq \delta_s, \quad r,s \in \{2,3,\dots,g\},
\end{equation}
where the null hypothesis $H_0$ implies that all treatment groups exhibit the same degree of difference from the control group, suggesting that stratifying patients by treatment group confers no additional benefit. The alternative hypothesis $H_1$ indicates that at least one group differs from the others, warranting differentiated treatment strategies across groups. 

\subsection{Testing Methods}\label{Sec4.1}
\subsubsection{Posterior Range Statistic}
Within the Bayesian framework, we propose a posterior range method for our hypothesis testing. For the $t$-th posterior sample, we compute the range of all risk differences, which is defined as the test statistic in (\ref{Define_W}):
\begin{equation}\label{Define_W}
	W^{(t)} = \max_{i=2,\dots,g} \delta_i^{(t)} - \min_{i=2,\dots,g} \delta_i^{(t)},
\end{equation}
where $\delta_i^{(t)}$ denotes the $t$-th posterior draw of the risk difference for group $i$. If all $\delta_i$ are equal, then $W = 0$; otherwise, $W > 0$. Since the range $W$ is a non-negative parameter whose posterior distribution is typically right-skewed and for which the point mass at zero is zero under a continuous posterior, testing whether $W$ is exactly zero is not statistically meaningful (Rouder et al., 2009\cite{rouder2009bayesian}). This testing method combines the posterior distribution of the range $W$ with the HDI+ROPE decision rule proposed by Kruschke (2018) \cite{kruschke2018rejecting}. The HDI+ROPE decision rule uses the HDI of the posterior distribution in conjunction with a ROPE around the null value. The HDI summarizes the range of most credible values of the parameter, while the ROPE defines an interval of parameter values that are practically equivalent to the null value for the specific application. This approach recognizes that in continuous parameter spaces, testing whether a parameter is exactly zero is statistically ill-posed. The ROPE approach defines a small interval around the null value that contains parameter values considered practically equivalent to zero for the specific application. In our study, we define the ROPE as $[0, \delta_0]$ since \(W\) is non-negative, where $\delta_0$ is the calibrated equivalence margin. Therefore, we reject $H_0$ if the $95\%$ HPD interval for $R$ lies entirely above $\delta_0$.

Let $[L_W, U_W]$ denote the $95\%$ HPD interval for $W$, defined as the shortest interval satisfying
\begin{equation}
	P(L_W \leq R \leq U_W \mid \boldsymbol{D}) = 0.95.
\end{equation}
Then,  we reject \(H_0\) if \(L_W > \delta_0\). That is, we reject the null hypothesis of equal risk differences only when the posterior evidence indicates that the range exceeds the clinically meaningful threshold with high probability. The choice of the equivalence margin $\delta_0$ should ideally be guided by clinical considerations. However, in the absence of a universally accepted threshold for bilateral data, we adopt a calibration approach proposed by Linde et al. (2023) \cite{linde2023decisions}. We calibrate $\delta_0$ by simulating data under both the null and alternative hypotheses, selecting a value that controls the empirical Type I error rate while maintaining reasonable empirical power.

\subsubsection{Bayes Factor}
To quantify the relative evidence for the null and alternative hypotheses, we compute the Bayes factor. Traditionally, the Bayes factor is computed as the ratio of marginal likelihoods under \(H_{0}\) and \(H_{1}\):
\begin{equation}
	BF_{01} = \frac{P(\boldsymbol{D} \mid H_0)}{P(\boldsymbol{D} \mid H_1)}.
\end{equation}
The Bayes factor $BF_{01} > 1$ indicates support for $H_0$, while $BF_{01} < 1$ favors $H_1$. Following Jeffreys (1961) \cite{jeffreys1961theory}, $BF_{01} > 3$ constitutes positive evidence for $H_0$, and $BF_{01} > 10$ constitutes strong evidence. However, computing marginal likelihoods is challenging. In addition, the approach to the prior is highly sensitive. For our nested models, the Bayes factor can be efficiently computed using the Savage-Dickey density ratio (Savage, 1962\cite{savage1962foundations}; Dickey, 1971 \cite{dickey1971weighted}). Recently, Barto{\v{s}} and Wagenmakers (2023)\cite{bartos2023general} developed an approximation for nested Bayes factors with informed priors that relies only on the maximum likelihood estimate and its standard error, and is computed via the Savage–Dickey density ratio. This approach is particularly relevant to our Savage--Dickey-based hypothesis testing. The Savage--Dickey ratio is given by
\begin{equation}
	BF_{01} = \frac{p(\delta_2 = 0, \delta_3 = 0, \dots, \delta_g = 0 \mid \boldsymbol{D})}
	{\pi(\delta_2 = 0, \delta_3 = 0, \dots, \delta_g = 0)},
\end{equation}
where the numerator is the joint posterior density evaluated at the null point, estimated via multivariate kernel density estimation from the MCMC samples. The denominator is the joint prior density at the null point, which can be obtained analytically.

\subsection{Testing Equality of Correlation Parameters}\label{Sec4.2}
A fundamental assumption in our model is that the correlation parameter $\gamma$ is common across all groups. To assess the validity of this assumption, we consider the additional hypothesis test:
\begin{equation}
	H_0^*: \gamma_1 = \gamma_2 = \dots = \gamma_g, \quad H_1^*: \exists\, r \neq s,\; \gamma_r \neq \gamma_s, \quad r,s = 2,3,\dots,g.
\end{equation}

We apply the same posterior range method. Let $W_\gamma = \max_i \gamma_i - \min_i \gamma_i$ denote the range of the group-specific correlation parameters, and let $[L_{W_\gamma}, U_{W_\gamma}]$ be its $95\%$ HPD interval. We reject $H_0^*$ if $L_{W_\gamma} > \delta_0$. The Bayes factor is
\begin{equation}
	BF_{01}^* = \frac{p(\gamma_1 = \gamma_2 = \dots = \gamma_g \mid \boldsymbol{D})}
	{\pi(\gamma_1 = \gamma_2 = \dots = \gamma_g)}.
\end{equation}
This test determines whether a saturated model is preferred over the reduced model with a common $\gamma$.

\subsection{Evaluation Criteria}\label{Sec4.3}
To evaluate the performance of the proposed testing procedure under finite sample conditions, we conduct Monte Carlo simulations and compute the following criteria.

(1) Under \(H_0\): $\delta_2 = \delta_3 = \dots = \delta_g$ (i.e., all risk differences are equal), the empirical type I error (TIE) rate is defined as the proportion of simulations in which the null hypothesis is incorrectly rejected:
\begin{equation}
	\text{TIE} = \frac{1}{N} \sum_{n=1}^{N} \mathbf{1}\big( L_W^{(n)} > \delta_0 \big),
\end{equation}
where $L_W^{(n)}$ is the lower bound of the HPD interval for the range in the $n$th simulated dataset, and $N$ is the number of replications.

(2) Under \(H_1\): at least one risk difference differs from the others (e.g., $\delta_2 = 0, \delta_3 = 0.2$), the empirical power is defined as the proportion of simulations in which the null hypothesis is correctly rejected:
\begin{equation}
	\text{Power} = \frac{1}{N} \sum_{n=1}^{N} \mathbf{1}\big( L_W^{(n)} > \delta_0 \big).
\end{equation}

The nominal significance level $\alpha = 0.05$ corresponds to the $95\%$ HPD interval used in the decision rule. A type I error rate close to $5\%$ indicates that the testing procedure is well-calibrated, while high power indicates sensitivity in detecting true differences (Li et al., 2023\cite{li2023testing}).

\section{Simulation studies}\label{Sec5}
In this section, we evaluate the performance of the proposed Bayesian methods, and compare them with the frequentist Wald method. We assess the interval estimation properties of the proposed Bayesian methods. Then, we calculate the TIEs and powers to investigate the performance of the proposed Bayesian testing procedure for the homogeneity of risk differences. All simulation experiments are implemented in Python (version 3.11.14).

\subsection{Comparisons of Bayesian and Frequentist Intervals}\label{Sec5.1}
In this section, we compare the performance of frequentist confidence intervals (FCIs) and the Bayesian credible intervals (BCIs) through three criteria. We use FCIs (Wald FCIs) or HPD BCIs since they have real coverage close to the nominal value. We also consider the estimated interval length and the mean squared error of the parameter estimates. To assess the accuracy of the estimators and interval estimates, we compute three criteria for various parameter combinations: Expected true coverage probability (ETCP), the expected width of credible intervals or confidence intervals (EWCI), and the expected mean squared error (EMSE). Furthermore, we calculate the Mean Marginal ETCP (METCP) as the average marginal coverage probability across all risk differences. 

\begin{figure}[htbp]
	\centering
	\begin{minipage}{\textwidth}
		\centering
		\includegraphics[width=0.95\textwidth]{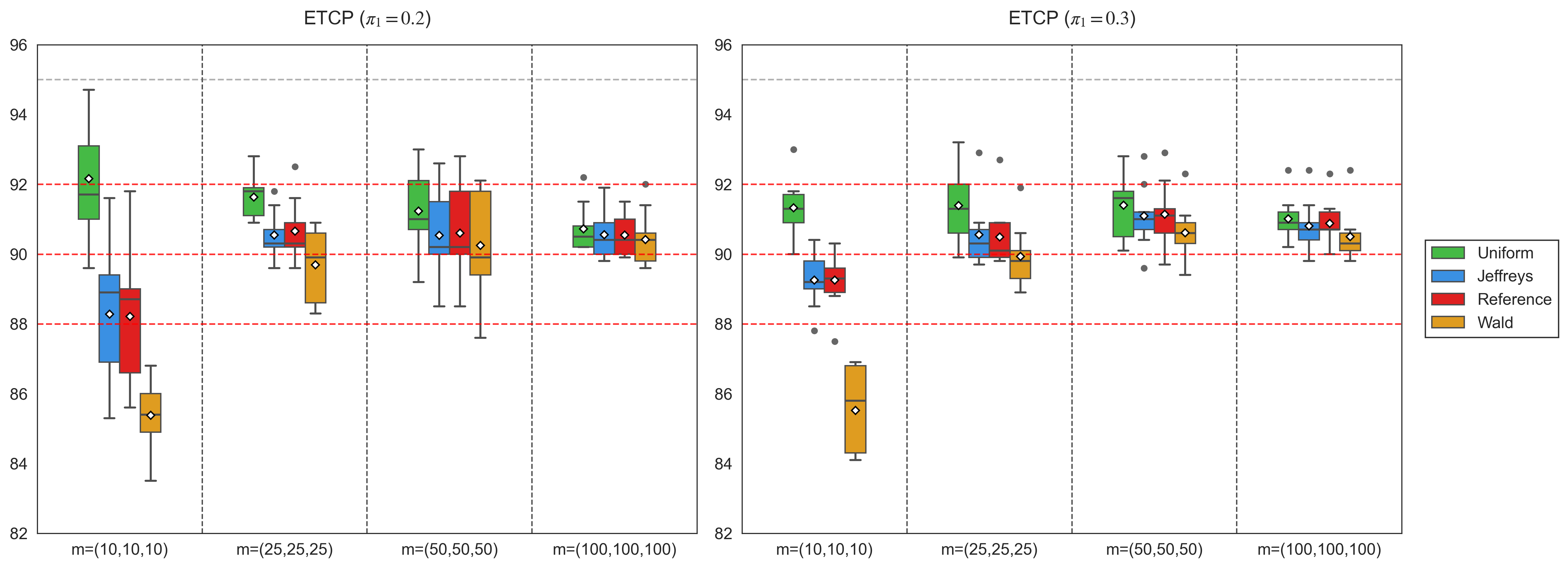}
	\end{minipage}
	\vspace{0.3cm}
	\begin{minipage}{\textwidth}
		\centering
		\includegraphics[width=0.95\textwidth]{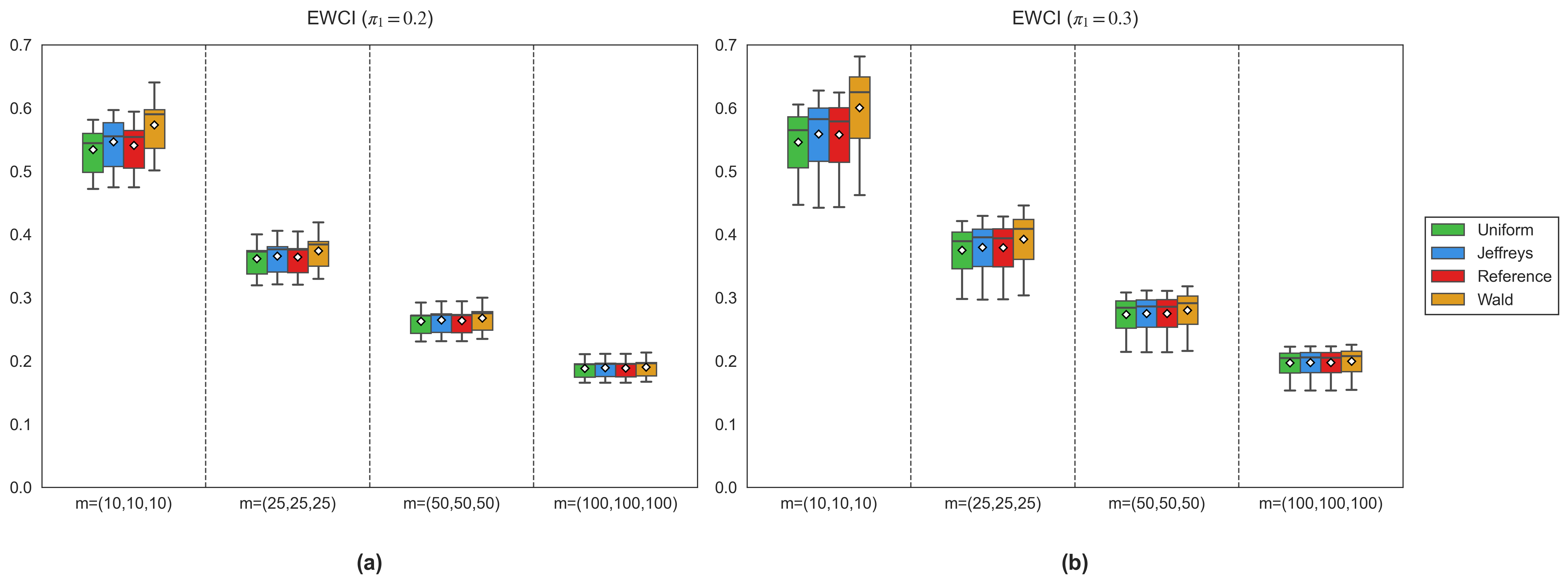}
	\end{minipage}
	\caption{Boxplots of ETCPs(\%) and EWCIs for various methods under balanced sample size for \(g=3\) with two baseline response rates: (a) \(\pi_1=0.2\) and (b) \(\pi_1=0.3\).}\label{fig:Boxplot_ETCP_EWCI_pi0.2-0.3_g3}
\end{figure}

\begin{figure}
	\centering
	\includegraphics[width=0.9\textwidth]{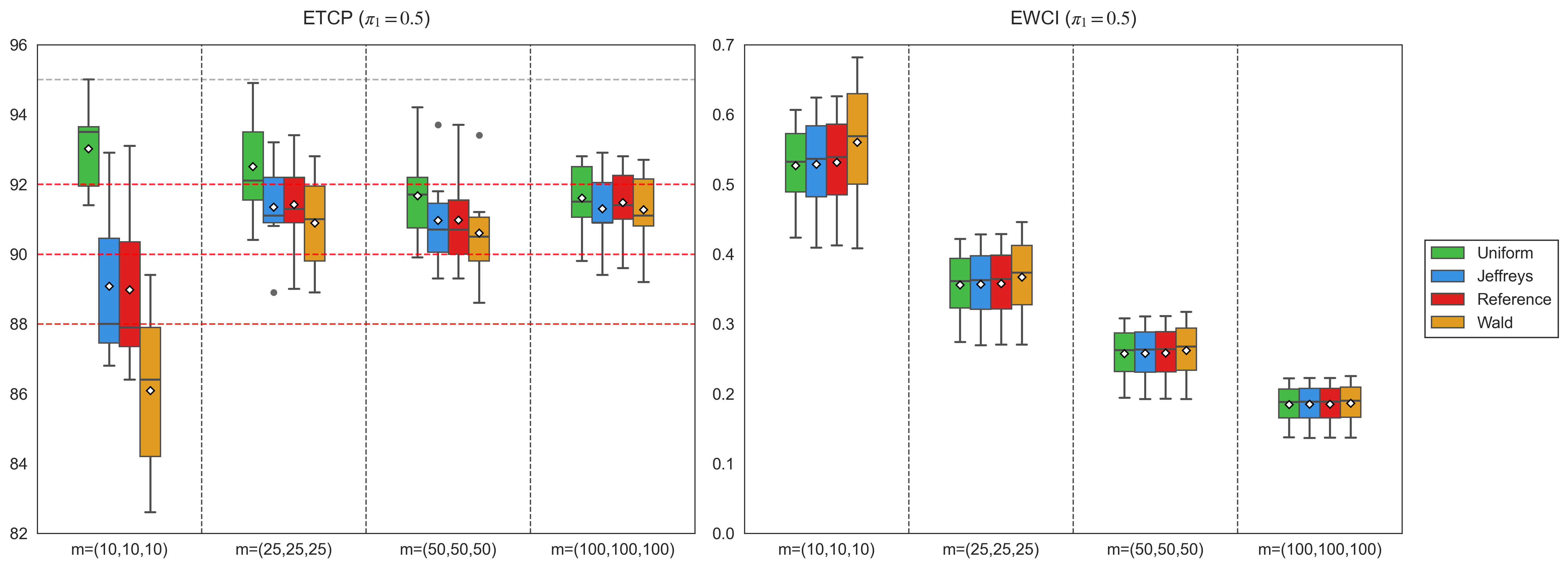}
	\caption{Boxplots of ETCPs(\%) and EWCIs for various methods under balanced sample size for \(g=3\) with \(\pi_1=0.5\).}\label{fig:Boxplot_ETCP_EWCI_pi0.5_g3}
\end{figure}

\begin{figure}
	\centering
	\includegraphics[width=0.9\textwidth]{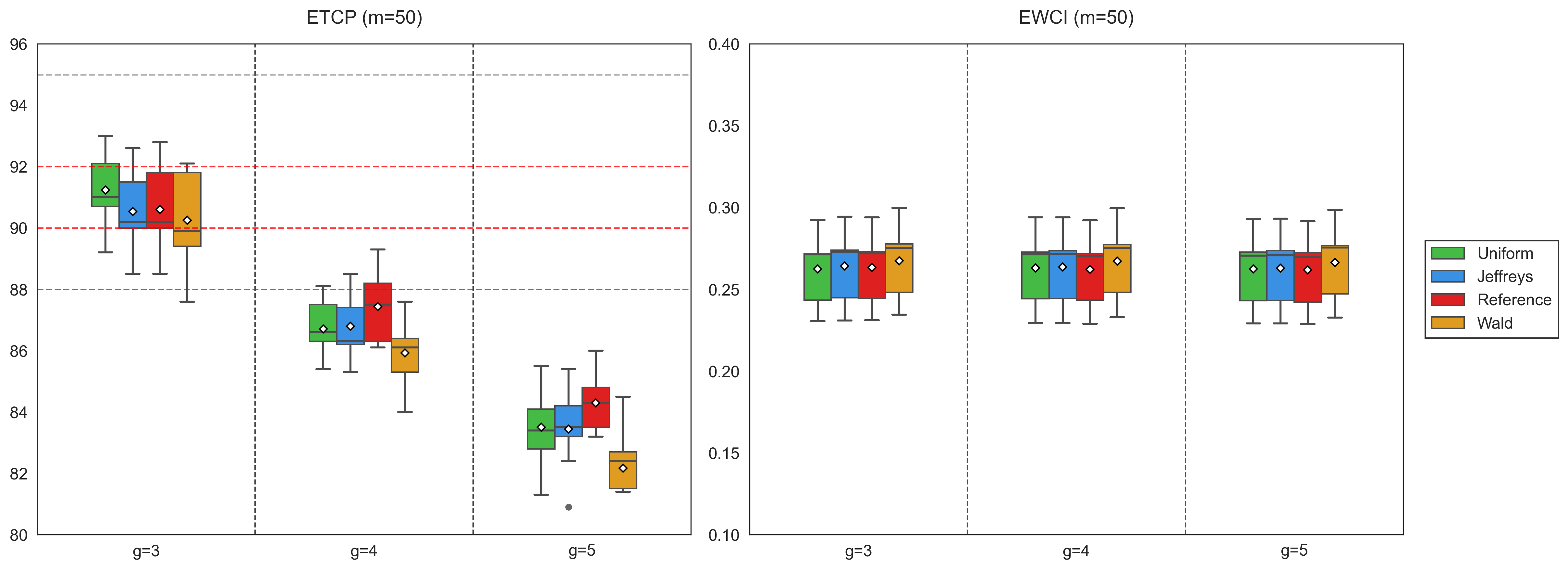}
	\caption{Boxplots of ETCPs(\%) and EWCIs for \(m=50\) with the groups \(g=3,4,5\).}\label{fig:Boxplot_ETCP_EWCI_g_m50}
\end{figure}

\begin{figure}
	\centering
	\includegraphics[width=0.9\textwidth]{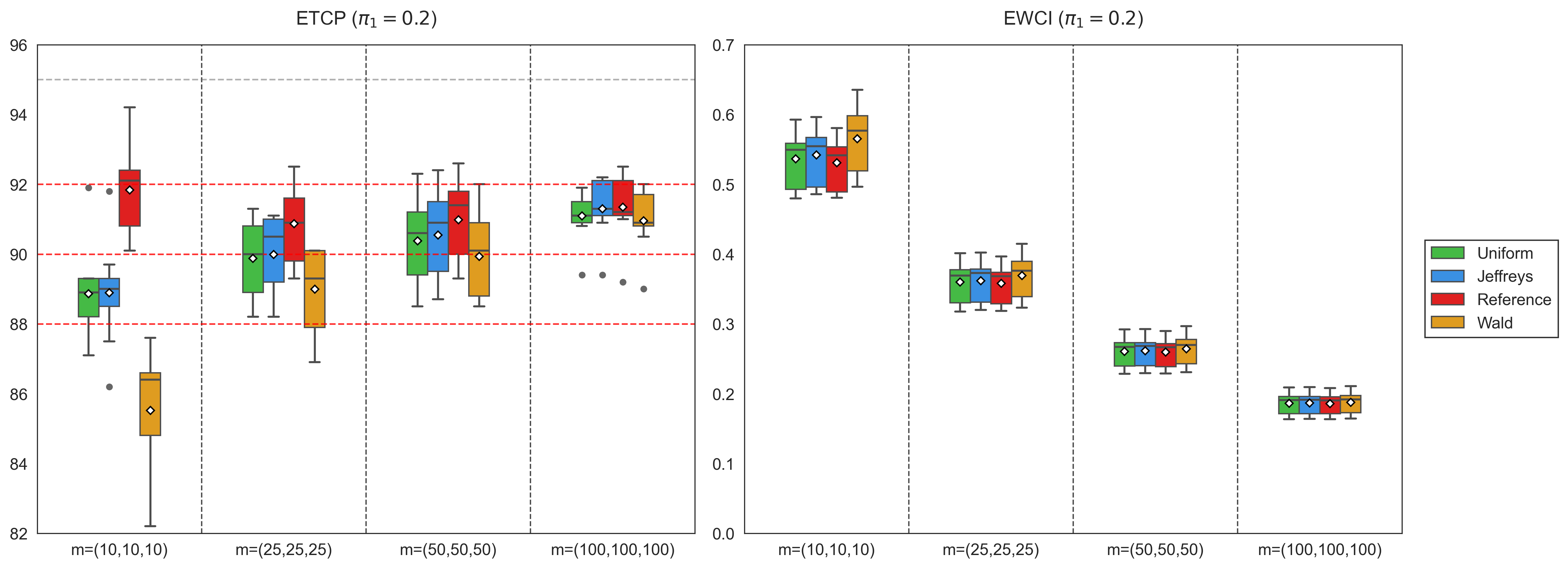}
	\caption{Boxplots of ETCPs(\%) and EWCIs for various methods under balanced sample size for \(g=3\) with \(\pi_1=0.5\).}\label{fig:Boxplot_ETCP_EWCI_pi0.2_unequal}
\end{figure}

Denote \(\boldsymbol{m}=(m_1,m_2,\dots,m_g)\), and \(\pi_1\) is the response rate of the control group. The parameter configurations studied are shown in Table \ref{tab:sim_param_config}. For the group design \(g=3\), we conduct Monte Carlo simulations with four equal sample sizes, i.e., \(m_1=m_2=\dots=m_g=m=10,25,50,100\). The boxplots of ETCP and EWCI of four methods are shown in Figures \ref{fig:Boxplot_ETCP_EWCI_pi0.2-0.3_g3} and   \ref{fig:Boxplot_ETCP_EWCI_pi0.5_g3}.
For \(m_1=m_2=\dots=m_g=50\), the groups \( g = 3, 4, 5\) are studied. We replicate each parameter combination independently 1,000 times. The simulation results are reported in
Tables \ref{tab:ci_performance_balanced} for balanced sample sizes with \(\pi_{1}=0.2\). Figure \ref{fig:Boxplot_ETCP_EWCI_g_m50} reports the boxplots of ETCP and EWCI of four methods for \( g = 3, 4, 5\). For each generated bilateral dataset, the Bayesian methods draw 10,000 samples from the posterior distribution, with the first 2,000 as burn-in. Then, we consider the three unequal risk difference configurations for \(g=3\): \(\boldsymbol{\delta_1}=(0.0, 0.2)\), \(\boldsymbol{\delta_2}=(0.1, 0.3)\) and \(\boldsymbol{\delta_3}=(-0.1, 0.1)\). The simulation results are reported in
Table \ref{tab:ci_performance_unequal_scenarios} for \(\pi_{1}=0.2\). The boxplots of ETCP and EWCI are reported in Figure \ref{fig:Boxplot_ETCP_EWCI_pi0.2_unequal}.
\begin{table}[htbp]
	\centering
	\caption{The parameter configurations for simulation studies.}
	\label{tab:sim_param_config}
	\begin{tabular*}{\textwidth}{@{\extracolsep\fill}llccc@{\extracolsep\fill}}
		\toprule
		\multirow{2}{*}{Parameter} &
		\multirow{2}{*}{Cases} &
		\multicolumn{3}{c}{Number of groups} \\
		\cmidrule(lr){3-5}
		& & \textbf{$g=3$} & \textbf{$g=4$} & \textbf{$g=5$} \\
		\midrule
		\multirow{3}{*}{$\pi_1$} 
		& I & $ 0.2 $ & $ 0.2 $ & $ 0.2 $\\
		& II & $ 0.3 $ & $ 0.3 $ & $ 0.3 $\\
		& III & $ 0.5 $ & $ 0.5 $ & $ 0.5 $\\
		\multirow{3}{*}{$\gamma$} 
		& $a_1$ & $ 0.2 $ & $ 0.2 $ & $ 0.2 $\\
		& $a_2$ & $ 0.3 $ & $ 0.3 $ & $ 0.3 $\\
		& $a_3$ & $ 0.5 $ & $ 0.5 $ & $ 0.5 $\\
		\multirow{3}{*}{$\delta$} 
		& $\delta_1$ & $ 0.0 $ & $ 0.0 $ & $ 0.0 $\\
		& $\delta_2$ & $ 0.1 $ & $ 0.1 $ & $ 0.1 $\\
		& $\delta_3$ & $ 0.3 $ & $ 0.3 $ & $ 0.3 $\\
		\bottomrule
	\end{tabular*}
\end{table}

\begin{sidewaystable}[htbp]
	\centering
	\caption{Performance of 95\% HPDs/CIs for the common risk difference with \(g=3\).}
	\label{tab:ci_performance_balanced}
	\begin{tabular*}{\textwidth}{@{\extracolsep\fill}lccccccccccccccccccc@{\extracolsep\fill}}
		\toprule
		\multirow{2}{*}{} & \multirow{2}{*}{$\delta$} & \multirow{2}{*}{$\pi_1$} & \multirow{2}{*}{$\gamma$} &  \multicolumn{4}{c}{$\boldsymbol{m}=(10,10,10)$} & \multicolumn{4}{c}{$\boldsymbol{m}=(25,25,25)$} & \multicolumn{4}{c}{$\boldsymbol{m}=(50,50,50)$} & \multicolumn{4}{c}{$\boldsymbol{m}=(100,100,100)$} \\
		\cmidrule(lr){5-8} \cmidrule(lr){9-12} \cmidrule(lr){13-16} \cmidrule(lr){17-20}
		& & & & $\text{CI}_\text{U}$ & $\text{CI}_\text{J}$ & $\text{CI}_\text{R}$ & $\text{CI}_\text{W}$ & $\text{CI}_\text{U}$ & $\text{CI}_\text{J}$ & $\text{CI}_\text{R}$ & $\text{CI}_\text{W}$ & $\text{CI}_\text{U}$ & $\text{CI}_\text{J}$ & $\text{CI}_\text{R}$ & $\text{CI}_\text{W}$ & $\text{CI}_\text{U}$ & $\text{CI}_\text{J}$ & $\text{CI}_\text{R}$ & $\text{CI}_\text{W}$ \\
		\midrule
		\multirow{9}{*}{METCP} 
		& \multirow{3}{*}{0.0}
		& \multirow{3}{*}{I} & $a_1$ & 97.20 & 92.80 & 92.70 & 91.50 & 95.70 & 95.00 & 95.00 & 95.00 & 94.90 & 94.20 & 94.20 & 94.00 & 94.60 & 94.40 & 94.30 & 94.20 \\ 
		& & & $a_2$ & 96.20 & 91.90 & 92.10 & 90.70 & 95.10 & 94.30 & 94.30 & 94.20 & 95.40 & 94.90 & 94.80 & 94.80 & 94.90 & 94.80 & 95.10 & 94.80 \\ 
		& & & $a_3$ & 97.00 & 93.10 & 93.10 & 91.60 & 95.60 & 94.80 & 94.80 & 94.80 & 96.00 & 95.80 & 96.00 & 95.60 & 94.40 & 94.80 & 94.30 & 94.20 \\ 
		& \multirow{3}{*}{0.1}
		& \multirow{3}{*}{I} & $a_1$ 
		& 95.60 & 94.00 & 93.90 & 92.30 & 94.90 & 94.20 & 94.30 & 93.70 & 95.20 & 94.80 & 95.00 & 94.60 & 94.80 & 94.50 & 94.70 & 94.50 \\ 
		& & & $a_2$ 
		& 95.50 & 94.00 & 94.00 & 92.30 & 95.60 & 94.80 & 94.70 & 94.30 & 95.00 & 94.70 & 94.40 & 94.20 & 95.70 & 95.60 & 95.40 & 95.60 \\ 
		& & & $a_3$ 
		& 94.30 & 92.40 & 92.60 & 91.80 & 96.20 & 95.50 & 96.00 & 95.00 & 94.90 & 94.50 & 94.60 & 94.50 & 95.10 & 94.80 & 94.90 & 94.60 \\ 
		& \multirow{3}{*}{0.3}
		& \multirow{3}{*}{I} & $a_1$ 
		& 95.20 & 94.60 & 94.20 & 91.00 & 95.20 & 94.60 & 94.60 & 93.80 & 94.00 & 93.60 & 93.60 & 93.20 & 94.80 & 94.90 & 94.80 & 94.80 \\ 
		& & & $a_2$ 
		& 94.80 & 94.00 & 94.00 & 92.00 & 95.60 & 95.0 & 95.40 & 94.20 & 95.90 & 95.60 & 95.80 & 95.60 & 95.10 & 95.20 & 95.20 & 95.40 \\ 
		& & & $a_3$ 
		& 96.00 & 95.30 & 95.40 & 92.40 & 95.00 & 94.60 & 94.60 & 93.20 & 96.00 & 95.50 & 95.60 & 95.80 & 94.60 & 94.50 & 94.50 & 94.40 \\ 
		\multirow{9}{*}{ETCP} 
		& \multirow{3}{*}{0.0}
		& \multirow{3}{*}{I} & $a_1$ 
		& 94.70 & 86.90 & 86.60 & 85.30 & 92.10 & 90.70 & 90.90 & 90.80 & 91.00 & 89.80 & 90.00 & 89.40 & 90.50 & 90.00 & 90.00 & 89.80 \\ 
		& & & $a_2$ 
		& 93.10 & 85.30 & 85.60 & 83.50 & 91.40 & 90.00 & 90.10 & 89.90 & 91.30 & 90.40 & 90.00 & 90.30 & 90.60 & 90.40 & 91.00 & 90.40 \\ 
		& & & $a_3$ 
		& 94.50 & 87.50 & 87.50 & 84.90 & 91.90 & 90.60 & 90.50 & 90.60 & 93.00 & 92.60 & 92.80 & 92.10 & 90.20 & 90.40 & 89.90 & 89.70 \\ 
		& \multirow{3}{*}{0.1}
		& \multirow{3}{*}{I} & $a_1$ 
		& 91.70 & 88.90 & 88.70 & 86.00 & 90.90 & 89.60 & 89.60 & 88.50 & 90.90 & 90.00 & 90.30 & 89.70 & 90.20 & 89.80 & 90.00 & 89.80 \\ 
		& & & $a_2$ 
		& 91.60 & 89.40 & 89.20 & 86.60 & 91.80 & 90.30 & 90.30 & 89.90 & 90.70 & 90.20 & 89.80 & 89.30 & 92.20 & 91.90 & 91.50 & 92.00 \\ 
		& & & $a_3$ 
		& 89.60 & 86.30 & 86.50 & 85.40 & 92.80 & 91.80 & 92.50 & 90.90 & 90.70 & 90.00 & 90.20 & 89.90 & 91.50 & 91.00 & 91.10 & 90.60 \\ 
		& \multirow{3}{*}{0.3}
		& \multirow{3}{*}{I} & $a_1$ 
		& 91.00 & 89.70 & 89.00 & 84.00 & 91.10 & 90.20 & 90.20 & 88.60 & 89.20 & 88.50 & 88.50 & 87.60 & 90.30 & 90.60 & 90.40 & 90.40 \\ 
		& & & $a_2$ 
		& 90.50 & 88.90 & 89.00 & 85.90 & 91.80 & 91.40 & 91.60 & 89.70 & 92.10 & 91.80 & 92.00 & 91.80 & 90.80 & 90.90 & 91.00 & 91.40 \\ 
		& & & $a_3$ 
		& 92.70 & 91.60 & 91.80 & 86.80 & 90.90 & 90.30 & 90.20 & 88.30 & 92.20 & 91.50 & 91.80 & 92.10 & 90.20 & 90.00 & 90.00 & 89.60 \\ 
		\multirow{6}{*}{EWCI} 
		& \multirow{2}{*}{0.0}
		& \multirow{2}{*}{I} & $a_1$ 
		& 0.560 & 0.577 & 0.564 & 0.590 & 0.374 & 0.380 & 0.376 & 0.384 & 0.271 & 0.274 & 0.272 & 0.275 & 0.195 & 0.196 & 0.195 & 0.196 \\ 
		& & & $a_2$
		& 0.533 & 0.546 & 0.537 & 0.561 & 0.357 & 0.362 & 0.358 & 0.366 & 0.259 & 0.261 & 0.259 & 0.262 & 0.185 & 0.185 & 0.185 & 0.186 \\ 
		& & & $a_3$ 
		& 0.488 & 0.496 & 0.489 & 0.510 & 0.324 & 0.327 & 0.325 & 0.331 & 0.235 & 0.236 & 0.235 & 0.238 & 0.167 & 0.168 & 0.168 & 0.168 \\ 
		& \multirow{2}{*}{0.1}
		& \multirow{2}{*}{I} & $a_1$ 
		& 0.578 & 0.596 & 0.589 & 0.627 & 0.393 & 0.400 & 0.397 & 0.409 & 0.287 & 0.290 & 0.289 & 0.294 & 0.207 & 0.208 & 0.207 & 0.209 \\ 
		& & & $a_2$ 
		& 0.550 & 0.566 & 0.560 & 0.597 & 0.374 & 0.379 & 0.377 & 0.389 & 0.272 & 0.274 & 0.273 & 0.277 & 0.195 & 0.196 & 0.196 & 0.197 \\
		& & & $a_3$ 
		& 0.498 & 0.507 & 0.504 & 0.536 & 0.337 & 0.341 & 0.339 & 0.350 & 0.243 & 0.245 & 0.244 & 0.248 & 0.174 & 0.175 & 0.174 & 0.176 \\
		& \multirow{2}{*}{0.3}
		& \multirow{2}{*}{I} & $a_1$ 
		& 0.581 & 0.596 & 0.594 & 0.640 & 0.400 & 0.405 & 0.405 & 0.419 & 0.292 & 0.294 & 0.294 & 0.300 & 0.210 & 0.211 & 0.211 & 0.213 \\ 
		& & & $a_2$ 
		& 0.544 & 0.555 & 0.554 & 0.594 & 0.372 & 0.376 & 0.376 & 0.388 & 0.271 & 0.273 & 0.273 & 0.278 & 0.194 & 0.195 & 0.195 & 0.197 \\ 
		& & & $a_3$ 
		& 0.472 & 0.474 & 0.474 & 0.501 & 0.320 & 0.321 & 0.321 & 0.330 & 0.231 & 0.231 & 0.231 & 0.234 & 0.165 & 0.165 & 0.166 & 0.167 \\ 
		\multirow{6}{*}{EMSE} 
		& \multirow{2}{*}{0.0}
		& \multirow{2}{*}{I} & $a_1$ 
		& 0.0172 & 0.0215 & 0.0204 & 0.0253 & 0.0084 & 0.0093 & 0.0091 & 0.0099 & 0.0046 & 0.0048 & 0.0047 & 0.0050 & 0.0025 & 0.0025 & 0.0025 & 0.0026 \\ 
		& & & $a_2$ 
		& 0.0166 & 0.0205 & 0.0196 & 0.0239 & 0.0080 & 0.0088 & 0.0086 & 0.0093 & 0.0042 & 0.0044 & 0.0043 & 0.0045 & 0.0022 & 0.0022 & 0.0022 & 0.0023 
		\\ 
		& & & $a_3$ 
		& 0.0133 & 0.0158 & 0.0153 & 0.0181 & 0.0067 & 0.0072 & 0.0071 & 0.0076 & 0.0033 & 0.0034 & 0.0034 & 0.0035 & 0.0018 & 0.0019 & 0.0019 & 0.0019 \\ 
		& \multirow{2}{*}{0.1}
		& \multirow{2}{*}{I} & $a_1$ 
		& 0.0199 & 0.0240 & 0.0233 & 0.0285 & 0.0099 & 0.0109 & 0.0107 & 0.0117 & 0.0053 & 0.0056 & 0.0055 & 0.0058 & 0.0027 & 0.0028 & 0.0028 & 0.0028 \\ 
		& & & $a_2$ 
		& 0.0177 & 0.0212 & 0.0207 & 0.0251 & 0.0088 & 0.0096 & 0.0095 & 0.0103 & 0.0048 & 0.0050 & 0.0050 & 0.0052 & 0.0023 & 0.0024 & 0.0024 & 0.0024 \\ 
		& & & $a_3$ 
		& 0.0167 & 0.0194 & 0.0191 & 0.0226 & 0.0068 & 0.0072 & 0.0072 & 0.0077 & 0.0040 & 0.0041 & 0.0041 & 0.0042 & 0.0020 & 0.0020 & 0.0020 & 0.0021 \\
		& \multirow{2}{*}{0.3}
		& \multirow{2}{*}{I} & $a_1$ 
		& 0.0233 & 0.0255 & 0.0254 & 0.0301 & 0.0105 & 0.0109 & 0.0109 & 0.0116 & 0.0061 & 0.0063 & 0.0063 & 0.0065 & 0.0029 & 0.0029 & 0.0029 & 0.0030 \\ 
		& & & $a_2$ 
		& 0.0207 & 0.0223 & 0.0223 & 0.0259 & 0.0089 & 0.0092 & 0.0092 & 0.0099 & 0.0045 & 0.0046 & 0.0046 & 0.0047 & 0.0024 & 0.0024 & 0.0024 & 0.0025 \\
		& & & $a_3$ 
		& 0.0147 & 0.0155 & 0.0155 & 0.0178 & 0.0067 & 0.0068 & 0.0068 & 0.0072 & 0.0033 & 0.0033 & 0.0033 & 0.0034 & 0.0018 & 0.0018 & 0.0018 & 0.0018 \\ 
		\bottomrule
	\end{tabular*}%
\end{sidewaystable}


\begin{sidewaystable}[htbp]
	\centering
	\caption{Performance of 95\% HPDs/CIs for the unequal risk difference configurations with \(g=3\).}
	\label{tab:ci_performance_unequal_scenarios}
	\begin{tabular*}{\textwidth}{@{\extracolsep\fill}lccccccccccccccccccc@{\extracolsep\fill}}
		\toprule
		\multirow{2}{*}{} & \multirow{2}{*}{$\delta$} & \multirow{2}{*}{$\pi_1$} & \multirow{2}{*}{$\gamma$} &  \multicolumn{4}{c}{$\boldsymbol{m}=(10,10,10)$} & \multicolumn{4}{c}{$\boldsymbol{m}=(25,25,25)$} & \multicolumn{4}{c}{$\boldsymbol{m}=(50,50,50)$} & \multicolumn{4}{c}{$\boldsymbol{m}=(100,100,100)$} \\
		\cmidrule(lr){5-8} \cmidrule(lr){9-12} \cmidrule(lr){13-16} \cmidrule(lr){17-20}
		& & & & $\text{CI}_\text{U}$ & $\text{CI}_\text{J}$ & $\text{CI}_\text{R}$ & $\text{CI}_\text{W}$ & $\text{CI}_\text{U}$ & $\text{CI}_\text{J}$ & $\text{CI}_\text{R}$ & $\text{CI}_\text{W}$ & $\text{CI}_\text{U}$ & $\text{CI}_\text{J}$ & $\text{CI}_\text{R}$ & $\text{CI}_\text{W}$ & $\text{CI}_\text{U}$ & $\text{CI}_\text{J}$ & $\text{CI}_\text{R}$ & $\text{CI}_\text{W}$ \\
		\midrule
		\multirow{9}{*}{METCP} 
		& \multirow{3}{*}{$\boldsymbol{\delta_1}$}
		& \multirow{3}{*}{I} & $a_1$ & 94.00 & 93.85 & 96.20 & 92.40 & 94.20 & 94.25 & 95.00 & 94.10 & 94.00 & 94.15 & 94.55 & 93.70 & 95.25 & 95.45 & 95.40 & 95.40 \\ 
		& & & $a_2$ & 92.90 & 93.05 & 95.70 & 91.40 & 94.40 & 94.55 & 94.90 & 94.25 & 95.80 & 95.80 & 96.00 & 95.55 & 95.55 & 95.70 & 95.90 & 95.45 \\ 
		& & & $a_3$ & 93.10 & 92.65 & 95.10 & 90.65 & 93.80 & 93.95 & 94.55 & 93.35 & 94.45 & 94.65 & 94.65 & 94.20 & 95.00 & 95.05 & 95.10 & 94.90 \\ 
		& \multirow{3}{*}{$\boldsymbol{\delta_2}$}
		& \multirow{3}{*}{I} & $a_1$ & 93.50 & 93.70 & 94.60 & 90.15 & 95.00 & 95.15 & 95.50 & 94.10 & 94.25 & 94.15 & 94.45 & 93.80 & 94.30 & 94.40 & 94.35 & 94.05 \\ 
		& & & $a_2$ & 93.60 & 93.65 & 95.25 & 92.30 & 93.55 & 93.75 & 93.85 & 93.20 & 95.20 & 95.40 & 95.60 & 94.90 & 94.95 & 95.10 & 94.90 & 94.85 \\ 
		& & & $a_3$ & 93.90 & 93.90 & 94.85 & 92.15 & 95.05 & 94.90 & 95.05 & 94.30 & 95.50 & 95.50 & 95.65 & 95.15 & 95.40 & 95.65 & 95.70 & 95.50 \\ 
		& \multirow{3}{*}{$\boldsymbol{\delta_3}$}
		& \multirow{3}{*}{I} & $a_1$ & 95.60 & 95.55 & 96.90 & 92.65 & 93.45 & 93.45 & 94.35 & 92.75 & 93.70 & 93.90 & 94.20 & 93.75 & 95.60 & 95.70 & 95.95 & 95.65 \\ 
		& & & $a_2$ & 94.15 & 94.45 & 95.95 & 92.00 & 94.90 & 95.00 & 95.70 & 94.30 & 94.80 & 94.95 & 95.30 & 94.45 & 94.90 & 94.85 & 95.05 & 94.90 \\ 
		& & & $a_3$ & 93.80 & 93.65 & 95.45 & 91.95 & 94.55 & 94.50 & 95.15 & 94.10 & 94.95 & 95.00 & 95.15 & 94.85 & 95.10 & 95.05 & 95.10 & 94.65 \\ 
		\multirow{9}{*}{ETCP} 
		& \multirow{3}{*}{$\boldsymbol{\delta_1}$}
		& \multirow{3}{*}{I} & $a_1$ & 89.30 & 89.00 & 92.90 & 86.20 & 89.40 & 89.50 & 90.80 & 89.30 & 89.40 & 89.50 & 90.50 & 88.80 & 90.90 & 91.30 & 91.20 & 91.10 \\ 
		& & & $a_2$ & 87.20 & 87.50 & 92.20 & 84.80 & 90.00 & 90.50 & 90.90 & 89.80 & 92.30 & 92.40 & 92.60 & 92.00 & 91.90 & 92.20 & 92.50 & 91.70 \\ 
		& & & $a_3$ & 87.10 & 86.20 & 90.80 & 82.60 & 88.40 & 88.70 & 89.80 & 87.70 & 89.30 & 89.70 & 89.70 & 88.80 & 91.10 & 91.10 & 91.30 & 90.90 \\ 
		& \multirow{3}{*}{$\boldsymbol{\delta_2}$}
		& \multirow{3}{*}{I} & $a_1$ & 88.20 & 88.50 & 90.10 & 82.20 & 90.80 & 91.00 & 91.60 & 89.00 & 89.40 & 89.20 & 90.00 & 88.50 & 89.40 & 89.40 & 89.20 & 89.00 \\ 
		& & & $a_2$ & 88.70 & 88.80 & 91.20 & 86.80 & 88.90 & 89.20 & 89.30 & 87.90 & 91.20 & 91.50 & 91.80 & 90.60 & 91.10 & 91.40 & 91.10 & 90.80 \\ 
		& & & $a_3$ & 88.90 & 89.30 & 90.60 & 86.40 & 91.30 & 91.00 & 91.10 & 90.10 & 91.80 & 91.90 & 92.10 & 91.20 & 91.50 & 92.10 & 92.10 & 91.80 \\ 
		& \multirow{3}{*}{$\boldsymbol{\delta_3}$}
		& \multirow{3}{*}{I} & $a_1$ & 91.90 & 91.80 & 94.20 & 87.60 & 88.20 & 88.20 & 89.80 & 86.90 & 88.50 & 88.70 & 89.30 & 88.50 & 91.90 & 92.10 & 92.50 & 92.00 \\ 
		& & & $a_2$ & 89.20 & 89.70 & 92.40 & 86.60 & 91.10 & 91.10 & 92.50 & 90.10 & 90.60 & 90.90 & 91.40 & 90.10 & 90.80 & 90.90 & 91.00 & 90.80 \\ 
		& & & $a_3$ & 89.30 & 89.20 & 92.10 & 86.50 & 90.80 & 90.70 & 92.00 & 90.10 & 90.90 & 91.10 & 91.40 & 90.90 & 91.20 & 91.20 & 91.20 & 90.50 \\ 
		\multirow{9}{*}{EWCI} 
		& \multirow{3}{*}{$\boldsymbol{\delta_1}$}
		& \multirow{3}{*}{I} & $a_1$ & 0.583 & 0.590 & 0.573 & 0.619 & 0.393 & 0.395 & 0.389 & 0.404 & 0.285 & 0.286 & 0.284 & 0.289 & 0.203 & 0.204 & 0.203 & 0.205 \\ 
		& & & $a_2$ & 0.549 & 0.554 & 0.542 & 0.582 & 0.371 & 0.373 & 0.368 & 0.381 & 0.269 & 0.269 & 0.267 & 0.273 & 0.192 & 0.192 & 0.192 & 0.194 \\ 
		& & & $a_3$ & 0.493 & 0.496 & 0.489 & 0.519 & 0.330 & 0.331 & 0.329 & 0.339 & 0.240 & 0.240 & 0.239 & 0.243 & 0.171 & 0.172 & 0.171 & 0.173 \\ 
		& \multirow{3}{*}{$\boldsymbol{\delta_2}$}
		& \multirow{3}{*}{I} & $a_1$ & 0.593 & 0.596 & 0.580 & 0.635 & 0.401 & 0.402 & 0.397 & 0.414 & 0.292 & 0.292 & 0.290 & 0.297 & 0.209 & 0.209 & 0.208 & 0.211 \\ 
		& & & $a_2$ & 0.559 & 0.562 & 0.549 & 0.598 & 0.378 & 0.378 & 0.374 & 0.390 & 0.273 & 0.273 & 0.272 & 0.277 & 0.196 & 0.196 & 0.195 & 0.198 \\ 
		& & & $a_3$ & 0.487 & 0.489 & 0.483 & 0.516 & 0.329 & 0.330 & 0.327 & 0.339 & 0.238 & 0.239 & 0.238 & 0.242 & 0.171 & 0.171 & 0.170 & 0.172 \\ 
		& \multirow{3}{*}{$\boldsymbol{\delta_3}$}
		& \multirow{3}{*}{I} & $a_1$ & 0.557 & 0.567 & 0.554 & 0.577 & 0.369 & 0.373 & 0.368 & 0.376 & 0.267 & 0.268 & 0.266 & 0.270 & 0.191 & 0.191 & 0.191 & 0.192 \\ 
		& & & $a_2$ & 0.527 & 0.536 & 0.526 & 0.544 & 0.350 & 0.353 & 0.350 & 0.357 & 0.254 & 0.255 & 0.254 & 0.256 & 0.180 & 0.181 & 0.180 & 0.181 \\ 
		& & & $a_3$ & 0.480 & 0.486 & 0.481 & 0.496 & 0.318 & 0.320 & 0.319 & 0.323 & 0.229 & 0.230 & 0.229 & 0.231 & 0.163 & 0.164 & 0.163 & 0.164 \\ 
		\multirow{9}{*}{EMSE} 
		& \multirow{3}{*}{$\boldsymbol{\delta_1}$}
		& \multirow{3}{*}{I} & $a_1$ & 0.0216 & 0.0222 & 0.0186 & 0.0264 & 0.0101 & 0.0102 & 0.0094 & 0.0110 & 0.0057 & 0.0057 & 0.0054 & 0.0059 & 0.0027 & 0.0027 & 0.0026 & 0.0027 \\ 
		& & & $a_2$ & 0.0215 & 0.0220 & 0.0187 & 0.0259 & 0.0095 & 0.0096 & 0.0089 & 0.0102 & 0.0045 & 0.0045 & 0.0044 & 0.0047 & 0.0024 & 0.0024 & 0.0023 & 0.0024 \\ 
		& & & $a_3$ & 0.0180 & 0.0182 & 0.0159 & 0.0211 & 0.0074 & 0.0074 & 0.0070 & 0.0079 & 0.0038 & 0.0038 & 0.0037 & 0.0039 & 0.0018 & 0.0018 & 0.0018 & 0.0019 \\ 
		& \multirow{3}{*}{$\boldsymbol{\delta_2}$}
		& \multirow{3}{*}{I} & $a_1$ & 0.0267 & 0.0271 & 0.0235 & 0.0323 & 0.0106 & 0.0107 & 0.0100 & 0.0115 & 0.0058 & 0.0059 & 0.0057 & 0.0061 & 0.0030 & 0.0030 & 0.0030 & 0.0031 \\ 
		& & & $a_2$ & 0.0226 & 0.0228 & 0.0202 & 0.0267 & 0.0105 & 0.0105 & 0.0100 & 0.0112 & 0.0050 & 0.0050 & 0.0048 & 0.0052 & 0.0025 & 0.0025 & 0.0025 & 0.0026 \\ 
		& & & $a_3$ & 0.0168 & 0.0169 & 0.0152 & 0.0196 & 0.0072 & 0.0072 & 0.0070 & 0.0076 & 0.0036 & 0.0036 & 0.0036 & 0.0038 & 0.0018 & 0.0018 & 0.0018 & 0.0018 \\ 
		& \multirow{3}{*}{$\boldsymbol{\delta_3}$}
		& \multirow{3}{*}{I} & $a_1$ & 0.0184 & 0.0192 & 0.0156 & 0.0229 & 0.0096 & 0.0099 & 0.0090 & 0.0105 & 0.0050 & 0.0051 & 0.0048 & 0.0052 & 0.0023 & 0.0024 & 0.0023 & 0.0024 \\ 
		& & & $a_2$ & 0.0181 & 0.0188 & 0.0156 & 0.0219 & 0.0079 & 0.0080 & 0.0074 & 0.0085 & 0.0043 & 0.0043 & 0.0041 & 0.0045 & 0.0021 & 0.0021 & 0.0021 & 0.0021 \\ 
		& & & $a_3$ & 0.0163 & 0.0168 & 0.0144 & 0.0192 & 0.0066 & 0.0067 & 0.0063 & 0.0071 & 0.0034 & 0.0035 & 0.0033 & 0.0036 & 0.0017 & 0.0017 & 0.0017 & 0.0018 \\ 
		\bottomrule
	\end{tabular*}%
\end{sidewaystable}

Table \ref{tab:ci_performance_balanced}, Table \ref{tab:ci_performance_unequal_scenarios} and Figures~\ref{fig:Boxplot_ETCP_EWCI_pi0.2-0.3_g3}--\ref{fig:Boxplot_ETCP_EWCI_pi0.2_unequal} summarize the interval estimation results. The findings for the ETCP criterion are summarized as follows.
\begin{enumerate}[label=\alph*.]
	\item As the sample size increases, the ETCP of all methods converges to the nominal level. In small samples, the Bayesian methods clearly outperform the Wald method, whose ETCP can fall below \(85\%\).
	\item As the number of groups increases from \(3\) to \(5\), the ETCP of all methods declines markedly, indicating that the multi-group setting degrades interval performance.
	\item Among the Bayesian priors, the uniform prior is the most conservative, while Bernardo's reference prior yields ETCP closest to the nominal level in the unequal-risk-difference scenario. The Jeffreys' and Bernardo's reference priors are nearly indistinguishable in most settings.
	\item At \(m=100\), all four methods yield similar ETCP values, and the differences among them become negligible.
\end{enumerate}

The findings for the EWCI criterion are as follows.
\begin{enumerate}[label=\alph*.]
	\item The EWCI of all methods decreases substantially as the sample size increases, with narrower boxplots indicating improved precision.
	\item The Wald method produces the widest intervals in small and moderate samples. Among the Bayesian methods, the uniform prior tends to yield the narrowest intervals, followed closely by the Jeffreys' and Bernardo's reference priors.
	\item In the unequal-risk-difference scenario, the Bernardo's reference prior produces the narrowest intervals, consistent with its superior ETCP performance.
	\item As the number of groups increases, the EWCI of all methods increases slightly. At \(m=100\), the differences among methods become negligible.
\end{enumerate}

The findings for the EMSE criterion are summarized as follows.
\begin{enumerate}[label=\alph*.]
	\item The EMSE of all methods decreases as the sample size increases. The Wald method has the highest EMSE in small samples, while the Bayesian methods perform substantially better.
	\item The uniform prior achieves the lowest EMSE in the equal-risk-difference and \(\pi_1=0.3\) scenarios, whereas the Bernardo's reference prior performs best in the unequal-risk-difference scenario.
	\item The Jeffreys' and Bernardo's reference priors behave similarly in most settings, with relative differences typically below \(10\%\).
	\item At \(m=100\), all four methods yield similar EMSE values, and the choice of method becomes less critical.
\end{enumerate}

In summary, the Bayesian methods outperformed the Wald method in small samples across all three criteria, while the differences among methods diminished as the sample size increased to \(m=100\).

\subsection{Performance of the Homogeneity Test}\label{Sec5.2}
In the simulation study, we calculate TIEs and powers to assess the performance of the proposed posterior range test for the homogeneity test of risk difference and compare it with the frequentist Wald test. We conduct the posterior range test under the three Bayesian priors and the frequentist Wald test at the significance level \(\alpha = 0.05\), with \(10{,}00\) replications randomly generated under the null or alternative hypothesis. All tests follow the settings in Section~\ref{Sec5.1} and Table~\ref{tab:sim_param_config}, but here we restrict attention to the case \(g=3\) and \(\pi_1=0.2\). Specifically, we use sample sizes \(m = 10, 25, 50, 100\) and correlation levels \(\gamma = 0.2, 0.3, 0.5\), with Group 1 as the reference group. First, the null hypothesis \(H_0: \delta_2 = \delta_3 = \cdots = \delta_g=\delta\) is evaluated under the risk difference \(\delta = 0, 0.1, 0.3\) (corresponding to \(\delta_1, \delta_2, \delta_3\) in Table~\ref{tab:sim_param_config}). Then, we calculate the empirical powers in the same parameter
configurations. The risk difference \(\boldsymbol{\delta_1}=(0.0, 0.2)\), \(\boldsymbol{\delta_2}=(0.1, 0.3)\) and \(\boldsymbol{\delta_3}=(-0.1, 0.1)\) are selected under the alternative
hypothesis \(H_1\). For the Bayesian tests, \(H_0\) is rejected if the 95\% HPD lower bound \(L_W\) of the posterior range \(W\) exceeds the calibrated equivalence margin \(\delta_0\).

\begin{table}[htbp]
	\centering
	\caption{Calibrated equivalence margins \(\delta_0\) for the posterior range test with \(g=3\).}
	\label{tab:calibration_g3}
	\begin{tabular*}{\textwidth}{@{\extracolsep\fill}lcccc@{\extracolsep\fill}}
		\toprule
		Prior & \(m=10\) & \(m=25\) & \(m=50\) & \(m=100\) \\
		\midrule
		Uniform   & 0.0070 & 0.0012 & 0.0045 & 0.0010 \\
		Jeffreys  & 0.0404 & 0.0102 & 0.0067 & 0.0020 \\
		Reference & 0.0170 & 0.0142 & 0.0032 & 0.0010 \\
		\bottomrule
	\end{tabular*}
	{\raggedright \small \textit{Notes:} Each value is calibrated via \(N=500\) simulated datasets under \(H_0:\delta_2=\delta_3\). The empirical TIE lies within \([0.044, 0.052]\) for all combinations.\par}
\end{table}

The equivalence margin \(\delta_0\) in the HDI+ROPE decision rule was calibrated separately for each combination of group number \(g\), sample size \(m\), and prior specification. For each \((g,m,\text{prior})\) combination, we simulated \(500\) datasets for calibration under the null hypothesis \(H_0:\delta_2=\cdots=\delta_g\), computed the 95\% HPD lower bound \(L_W\) of the posterior range \(W\), and selected the value of \(\delta_0\) that yielded an empirical TIE rate closest to the nominal \(5\%\) level. Table~\ref{tab:calibration_g3} reports the calibrated margins for \(g=3\), ranging from 0.0010 to 0.0404 across sample sizes and priors, with empirical TIEs within \([0.044,0.052]\). The calibrated \(\delta_0\) generally decreases with the sample size, reflecting the narrowing of the posterior range. The small non-monotonicity at intermediate sample sizes reflects the discreteness of the trinomial likelihood and the finite number of calibration datasets. Using these calibrated margins, we now assess the frequentist properties of the posterior range test for \(g=3\). The final reported TIEs and powers in Table~\ref{tab:TIEs} are evaluated with an independent simulation study using \(1000\) replications per scenario.

The results of Table~\ref{tab:TIEs} show that the four methods differ mainly in their TIE control and power. For the empirical TIEs, Bernardo's reference prior yields values closest to the nominal \(5\%\) level. The Jeffreys' prior is conservative, while the uniform prior is occasionally liberal. The Wald test has inflated TIEs at \(m=10\) and \(m=25\), ranging from \(6.6\%\) to \(9.3\%\) and from \(5.3\%\) to \(7.5\%\), respectively. At \(m=100\), all methods perform similarly. Power increases with sample size and the correlation parameter \(\gamma\). At small samples, the Wald test has higher power, but this is partly due to its inflated TIEs. The Bernardo's reference prior maintains competitive power while keeping the TIE near \(5\%\). At \(m=100\), powers approach \(100\%\), and the uniform and Bernardo's reference priors outperform the Jeffreys' prior. In summary, the Bernardo's reference prior provides the best balance between TIE control and power.

For \(g=4\) and \(g=5\), the same calibration procedure yields larger
margins because the posterior range widens with the number of groups. The
calibrated values are reported in Table~\ref{tab:calibration_appendix} in
Appendix~\ref{app:B}, and range from 0.037 to 0.138 for \(g=4\)
and from 0.056 to 0.228 for \(g=5\). Since a full simulation study for
\(g\ge 4\) requires substantially more computation and is beyond the scope
of the present paper, we focus on \(g=3\) in the main simulation and leave
a systematic investigation of \(g\ge 4\) to future work.
\begin{table}[htbp]
	\centering
	\caption{The empirical TIEs and powers (\%) for $g = 3$ with \(\pi_1=0.2\)}
	\label{tab:TIEs}
	\begin{tabular*}{\textwidth}{@{\extracolsep\fill}lccccccccccccccccccc@{\extracolsep\fill}}
		\toprule
		\multirow{2}{*}{groups} & \multirow{2}{*}{$\delta$} & \multirow{2}{*}{$\pi_1$} & \multirow{2}{*}{$\gamma$} &  \multicolumn{4}{c}{$\boldsymbol{m}=(10,10,10)$} & \multicolumn{4}{c}{$\boldsymbol{m}=(25,25,25)$} & \multicolumn{4}{c}{$\boldsymbol{m}=(50,50,50)$} & \multicolumn{4}{c}{$\boldsymbol{m}=(100,100,100)$} \\
		\cmidrule(lr){5-8} \cmidrule(lr){9-12} \cmidrule(lr){13-16} \cmidrule(lr){17-20}
		& & & & $T_U$ & $T_J$ & $T_R$ & $T_W$ & $T_U$ & $T_J$ & $T_R$ & $T_W$ & $T_U$ & $T_J$ & $T_R$ & $T_W$ & $T_U$ & $T_J$ & $T_R$ & $T_W$ \\
		\midrule
		\multirow{18}{*}{$g=3$} 
		& \multirow{3}{*}{0.0}
		& \multirow{3}{*}{I} 
		& $a_1$ & 4.20 & 3.70 & 5.30 & 9.10 & 5.20 & 3.40 & 3.50 & 7.50 & 3.90 & 4.10 & 4.80 & 6.50 & 5.00 & 5.10 & 5.80 & 5.80 \\ 
		& & & $a_2$ & 3.70 & 3.00 & 4.00 & 8.60 & 6.20 & 3.80 & 3.10 & 6.00 & 3.00 & 2.90 & 4.10 & 5.10 & 4.70 & 4.20 & 5.20 & 4.60 \\ 
		& & & $a_3$ & 3.50 & 2.70 & 3.30 & 7.60 & 4.50 & 2.80 & 2.70 & 7.10 & 3.80 & 3.50 & 4.60 & 3.90 & 4.90 & 4.00 & 5.20 & 5.60 \\ 
		& \multirow{3}{*}{0.1}
		& \multirow{3}{*}{I} 
		& $a_1$ & 4.90 & 3.60 & 4.60 & 6.60 & 6.60 & 4.50 & 3.80 & 6.50 & 4.20 & 3.90 & 4.60 & 5.70 & 5.20 & 5.00 & 5.70 & 6.30 \\ 
		& & & $a_2$ & 5.80 & 5.10 & 6.10 & 8.00 & 7.30 & 4.50 & 3.70 & 5.30 & 4.60 & 4.50 & 5.70 & 5.60 & 5.90 & 5.40 & 6.80 & 4.80 \\ 
		& & & $a_3$ & 4.40 & 3.70 & 4.00 & 9.30 & 5.30 & 3.60 & 2.90 & 5.80 & 4.00 & 3.60 & 4.50 & 5.70 & 6.60 & 5.10 & 6.10 & 5.30 \\ 
		& \multirow{3}{*}{0.3}
		& \multirow{3}{*}{I} 
		& $a_1$ & 5.90 & 4.60 & 5.40 & 7.40 & 7.10 & 4.70 & 3.90 & 7.10 & 5.90 & 5.10 & 6.60 & 5.00 & 3.70 & 4.10 & 4.40 & 5.30 \\ 
		& & & $a_2$ & 4.50 & 3.30 & 3.90 & 7.60 & 5.30 & 3.80 & 3.10 & 5.90 & 5.20 & 5.20 & 6.00 & 6.30 & 6.30 & 5.20 & 7.00 & 4.70 \\ 
		& & & $a_3$ & 4.10 & 3.30 & 4.70 & 9.00 & 6.10 & 4.70 & 4.70 & 6.10 & 4.10 & 4.50 & 5.80 & 6.80 & 5.40 & 4.50 & 5.50 & 5.80 \\  
		& \multirow{3}{*}{$\boldsymbol{\delta_1}$}
		& \multirow{3}{*}{I} 
		& $a_1$ & 17.7 & 17.0 & 19.4 & 24.9 & 46.1 & 40.9 & 38.0 & 42.0 & 69.3 & 68.6 & 70.9 & 63.3 & 95.9 & 95.9 & 96.7 & 94.2 \\ 
		& & & $a_2$ & 20.4 & 19.3 & 20.5 & 26.2 & 52.4 & 45.8 & 45.1 & 46.6 & 74.0 & 73.1 & 76.4 & 69.2 & 97.3 & 96.9 & 97.8 & 94.9 \\ 
		& & & $a_3$ & 27.8 & 26.5 & 27.4 & 29.1 & 62.9 & 55.2 & 52.8 & 53.2 & 82.6 & 81.8 & 83.4 & 80.8 & 98.9 & 98.7 & 99.2 & 97.1 \\ 
		& \multirow{3}{*}{$\boldsymbol{\delta_2}$}
		& \multirow{3}{*}{I} & $a_1$ 
		& 20.3 & 18.9 & 20.5 & 24.5 & 41.4 & 35.4 & 32.8 & 40.1 & 65.6 & 64.3 & 67.4 & 62.3 & 93.5 & 92.9 & 93.8 & 89.3 \\ 
		& & & $a_2$ 
		& 20.6 & 19.7 & 20.7 & 24.9 & 52.6 & 44.9 & 43.7 & 43.4 & 74.2 & 74.2 & 75.4 & 66.6 & 95.3 & 94.7 & 95.5 & 92.1 \\ 
		& & & $a_3$ 
		& 27.0 & 25.3 & 27.2 & 29.5 & 62.0 & 57.3 & 54.6 & 54.0 & 85.9 & 86.0 & 86.8 & 79.1 & 99.4 & 99.2 & 99.4 & 97.6 \\ 
		& \multirow{3}{*}{$\boldsymbol{\delta_3}$}
		& \multirow{3}{*}{I} & $a_1$ 
		& 19.0 & 18.4 & 21.2 & 32.1 & 54.9 & 50.3 & 47.0 & 52.0 & 79.9 & 80.0 & 81.7 & 80.1 & 98.6 & 98.7 & 99.0 & 96.9 \\ 
		& & & $a_2$ 
		& 22.5 & 21.8 & 24.0 & 31.0 & 58.5 & 53.9 & 52.7 & 56.4 & 82.9 & 83.2 & 85.0 & 84.1 & 99.4 & 99.2 & 99.3 & 97.8 \\ 
		& & & $a_3$ 
		& 27.1 & 25.4 & 28.2 & 35.7 & 68.9 & 63.3 & 60.0 & 61.1 & 91.0 & 90.2 & 91.6 & 90.8 & 100.0 & 100.0 & 100.0 & 99.4 \\  
		\bottomrule
	\end{tabular*}
\end{table}

\section{Real examples}\label{Sec6}
In this section, we consider two real examples to illustrate the proposed methods. 

\textit{Example 1}. 
We illustrate the proposed methodology using the scleroderma clinical trial data previously analyzed by Postlethwaite et al. (2008)\cite{Postlethwaite2008}. In this double-blind, placebo-controlled trial, 168 patients with diffuse cutaneous systemic sclerosis were randomized to receive either oral type I collagen (\(n=46\)) or placebo (\(n=61\)) for the forearms. For each patient, the improvement status of both forearms was recorded at month 15, where a forearm was considered improved if the modified Rodnan Skin Score (MRSS) decreased by at least 2 units or became zero. Table~\ref{tab:scleroderma_data} summarizes the number of patients with 0, 1, or 2 improved forearms in each treatment group.

The primary objective is to compare the improvement rates between the collagen and placebo groups and to assess whether they share the same correlation structure. Since our framework reduces to Dallal's model when \(g=2\). We apply uniform, Jeffreys', and reference priors; the original analysis reported only the latter two, so we focus on those. Posterior sampling follows the scheme in Section~\ref{Sec3.4}. For each prior, we report posterior means, standard deviations, and 95\% HPD intervals for the risk difference \(\delta=\pi_2-\pi_1\), risk ratio \(R=\pi_2/\pi_1\), odds ratio \(\psi\), and correlation parameter \(\gamma\). We test \(H_0:\pi_1=\pi_2\) and \(H_0^*:\gamma_1=\gamma_2\) using posterior probabilities and Savage--Dickey Bayes factors, where \(H_0^*\) is tested via the saturated model. Because the Bayes factor supports the reduced model, we base posterior summaries on a common \(\gamma\). We compare our estimates of \(R\), \(\delta\), and \(\gamma\), and the Bayes factors for \(H_0\) and \(H_0^*\), with those in M'lan and Chen (2015) to confirm the correctness of our extension.

Convergence diagnostics based on four parallel chains yield \(\widehat{R}=1.0000\) for all priors, and the Metropolis--Hastings acceptance rate under the uniform prior is \(0.486\). Table~\ref{tab:scleroderma_estimates} reports posterior summaries. The posterior means of \(R\) are \(2.4771\) (Reference), \(2.4602\) (Jeffreys'), and \(2.3694\) (Uniform), with 95\% HPD intervals \((0.6651, 5.0559)\), \((0.6400, 4.9583)\), and \((0.6467, 4.6893)\), respectively. These estimates closely match the values \(2.481\) and \(2.479\) reported by M'lan and Chen (2015) under the reference and Jeffreys' priors, confirming the correctness of our extension. The posterior means of \(\delta\) are \(0.0922\), \(0.0939\), and \(0.0918\), and all three 95\% HPD intervals contain zero, indicating no significant difference in improvement rates between the collagen and placebo groups. This is consistent with Pei et al.\ (2012)\cite{pei2012confidence}, who reported \(\hat{\lambda}=0.0970\) with a 95\% confidence interval \((-0.0214, 0.2217)\) under the equal correlation model. The posterior means of \(\psi\) are \(2.8418\), \(2.8309\), and \(2.7167\); as expected, the odds ratio exceeds the risk ratio, with wider HPD intervals. The posterior means of \(\gamma\) are approximately \(0.29\) across all priors, with 95\% HPD intervals approximately \((0.108, 0.485)\), indicating a moderate positive correlation between paired forearms.

Table~\ref{tab:scleroderma_tests} reports posterior probabilities and Bayes factors. The probabilities \(P(\delta>0\mid D)\), \(P(R>1\mid D)\), and \(P(\psi>1\mid D)\) are approximately \(0.955\) across all priors, suggesting that the collagen group has a higher improvement rate than the placebo group with about \(95.5\%\) posterior probability. However, the Savage--Dickey Bayes factors \(BF_{01}\) for \(H_0:\delta=0\) versus \(H_1:\delta>0\) are \(1.2493\) (Reference), \(1.2638\) (Jeffreys'), and \(1.4065\) (Uniform), providing only negligible evidence in favor of the null. Thus, the evidence for a difference in improvement rates is weak, and the null hypothesis of equal cure rates cannot be rejected. The three priors yield highly consistent results, with differences in posterior means typically below \(0.01\) for \(\delta\) and below \(0.1\) for \(R\), consistent with M'lan and Chen (2015)\cite{mlan2015objective}. Overall, the scleroderma example confirms that our multi-group Bayesian framework correctly reduces to Dallal's model when \(g=2\).
\begin{table}[htbp]
	\centering
	\caption{Number of scleroderma patients whose forearm MRSS decreased by 2 or 3, or has 0 MRSS at month 15.}
	\label{tab:scleroderma_data}
	\begin{tabular*}{\textwidth}{@{\extracolsep\fill}lcc@{\extracolsep\fill}}
		\toprule
		\multirow{2}{*}{Number of forearms with improvement} & \multicolumn{2}{c}{Treatment Group} \\
		\cline{2-3}
		& Collagen & Placebo \\
		\hline
		0 & 36 & 55 \\
		1 & 4 & 3 \\
		2 & 6 & 3 \\
		Total & 46 & 61 \\
		\bottomrule
	\end{tabular*}
\end{table}
\begin{table}[htbp]
	\centering
	\caption{Posterior means, standard deviations, and 95\% HPD intervals for the scleroderma data.}
	\label{tab:scleroderma_estimates}
	\begin{tabular*}{\textwidth}{@{\extracolsep\fill}lccccccccc@{\extracolsep\fill}}
		\toprule
		\multirow{2}{*}{Parameter} & \multicolumn{3}{c}{Reference} & \multicolumn{3}{c}{Jeffreys'} & \multicolumn{3}{c}{Uniform} \\
		\cmidrule(lr){2-4} \cmidrule(lr){5-7} \cmidrule(lr){8-10}
		& Mean & Std & 95\% HPD & Mean & Std & 95\% HPD & Mean & Std & 95\% HPD \\
		\midrule
		$\delta$  & 0.0922 & 0.0563 & $(-0.0164, 0.2052)$ & 0.0939 & 0.0572 & $(-0.0160, 0.2075)$ & 0.0918 & 0.0564 & $(-0.0156, 0.2046)$ \\
		$R$       & 2.4771 & 1.3399 & $(0.6651, 5.0559)$ & 2.4602 & 1.3356 & $(0.6400, 4.9583)$ & 2.3694 & 1.2143 & $(0.6467, 4.6893)$ \\
		$\psi$    & 2.8418 & 1.7388 & $(0.5830, 6.1232)$ & 2.8309 & 1.7422 & $(0.5361, 6.0243)$ & 2.7167 & 1.5827 & $(0.5896, 5.7276)$ \\
		$\gamma$  & 0.2907 & 0.0993 & $(0.1082, 0.4851)$ & 0.2907 & 0.0993 & $(0.1082, 0.4851)$ & 0.2941 & 0.0976 & $(0.1123, 0.4835)$ \\
		$\pi_1$   & 0.0817 & 0.0309 & $(0.0266, 0.1426)$ & 0.0840 & 0.0313 & $(0.0293, 0.1470)$ & 0.0863 & 0.0313 & $(0.0303, 0.1485)$ \\
		$\pi_2$   & 0.1739 & 0.0488 & $(0.0832, 0.2702)$ & 0.1779 & 0.0492 & $(0.0871, 0.2760)$ & 0.1781 & 0.0487 & $(0.0891, 0.2766)$ \\
		\bottomrule
	\end{tabular*}
\end{table}

\begin{table}[htbp]
	\centering
	\caption{Posterior probabilities and Bayes factors for the scleroderma data.}
	\label{tab:scleroderma_tests}
	\begin{tabular*}{\textwidth}{@{\extracolsep\fill}lccc@{\extracolsep\fill}}
		\toprule
		\multirow{2}{*}{Quantity} & \multicolumn{3}{c}{Prior} \\
		\cmidrule(lr){2-4}
		& Reference & Jeffreys' & Uniform \\
		\midrule
		$P(\delta>0\mid D)$ & 0.9554 & 0.9553 & 0.9539 \\
		$P(R>1\mid D)$      & 0.9554 & 0.9553 & 0.9539 \\
		$P(\psi>1\mid D)$   & 0.9554 & 0.9553 & 0.9539 \\
		$BF_{01}$           & 1.2493 & 1.2638 & 1.4065 \\
		\bottomrule
	\end{tabular*}
\end{table}

\textit{Example 2}.
We further illustrate the proposed methodology using the retinitis pigmentosa (RP) data from Berson et al.~(1980)\cite{Berson1980Retinitis}. The data consist of 216 patients aged 20--39 years, classified into four genetic groups: autosomal dominant RP (DOM, \(n=28\)), autosomal recessive RP (AR, \(n=21\)), sex-linked RP (SL, \(n=19\)), and isolated RP (ISO, \(n=148\)). For each patient, the number of affected eyes (0, 1, or 2) was recorded, giving a \(3 \times 4\) bilateral data structure summarized in Table~\ref{tab:rp_data}. This dataset is particularly suitable for our study because it involves more than two groups, uses an unbalanced design, and has moderate to large sample sizes across groups, allowing us to assess the proposed multi-group Bayesian framework in a realistic setting.

For each prior (uniform, Jeffreys', and reference), posterior samples are drawn as described in Section~\ref{Sec3.4}. Under Dallal's reduced model with a common \(\gamma\), we report posterior means, standard deviations, and 95\% HPD intervals for \(\pi_i\), \(\gamma\), and the effect measures relative to the reference group (DOM): the risk differences \(\delta_i=\pi_i-\pi_1\), risk ratios \(R_i=\pi_i/\pi_1\), and odds ratios \(\psi_i\), for \(i=2,3,4\). To test correlation homogeneity across the four groups, we fit the saturated model and apply the HDI+ROPE decision rule to the posterior range \(W=\max_i\gamma_i-\min_i\gamma_i\), using the average calibrated margins for \(g=4\) from Appendix~\ref{app:B}: \(\delta_0=0.0660\) (uniform), \(0.0810\) (Jeffreys'), and \(0.0740\) (reference). Since the RP data are unbalanced, \((m_1,\dots,m_4)=(28,21,19,148)\), no single calibrated value applies. For the risk differences, we test each \(H_{0i}:\delta_i=0\) individually using the Savage--Dickey density ratio and report \(BF_{01}\) and \(P(\delta_i>0\mid D)\). These analyses address two questions: whether the dependence structure is homogeneous across groups, and whether each group differs from the reference in its marginal response rate.

\begin{table}[!htbp]
	\centering
	\caption{The number of patients for genetic types.}
	\label{tab:rp_data}
	\begin{tabular*}{\textwidth}{@{\extracolsep\fill}lccccc@{\extracolsep\fill}}
		\toprule
		\multirow{2}{*}{Response} & \multicolumn{4}{c}{Genetic Group} & \multirow{2}{*}{Total} \\
		\cmidrule{2-5}
		& DOM & AR & SL & ISO & \\
		\midrule 
		0   & 15 & 7  & 3  & 67 & 92 \\
		1   & 6  & 5  & 2  & 24 & 37 \\
		2   & 7  & 9  & 14 & 57 & 87 \\
		Total & 28 & 21 & 19 & 148 & 216 \\
		\bottomrule
	\end{tabular*}
\end{table}
\begin{table}[htbp]
	\centering
	\caption{Posterior means and 95\% HPD intervals for the RP data under Dallal's reduced model with a common \(\gamma\).}
	\label{tab:rp_estimates}
	\begin{tabular*}{\textwidth}{@{\extracolsep\fill}lcccccc@{\extracolsep\fill}}
		\toprule
		\multirow{2}{*}{Parameter} & \multicolumn{2}{c}{Reference} & \multicolumn{2}{c}{Jeffreys'} & \multicolumn{2}{c}{Uniform} \\
		\cmidrule(lr){2-3} \cmidrule(lr){4-5} \cmidrule(lr){6-7}
		& Mean & 95\% HPD & Mean & 95\% HPD & Mean & 95\% HPD \\
		\midrule
		$\pi_{\text{DOM}}$ & 0.3957 & (0.2452, 0.5489) & 0.3969 & (0.2458, 0.5500) & 0.3970 & (0.2452, 0.5458) \\
		$\pi_{\text{AR}}$  & 0.5602 & (0.3916, 0.7193) & 0.5616 & (0.3977, 0.7254) & 0.5545 & (0.3875, 0.7116) \\
		$\pi_{\text{SL}}$  & 0.7013 & (0.5586, 0.8326) & 0.7013 & (0.5573, 0.8299) & 0.6885 & (0.5439, 0.8208) \\
		$\pi_{\text{ISO}}$ & 0.4651 & (0.3940, 0.5360) & 0.4657 & (0.3938, 0.5361) & 0.4649 & (0.3943, 0.5358) \\
		$\gamma$            & 0.1770 & (0.1236, 0.2343) & 0.1770 & (0.1236, 0.2343) & 0.1767 & (0.1252, 0.2333) \\
		$\delta_{\text{AR}}$  & 0.1645 & $(-0.0586, 0.3886)$ & 0.1648 & $(-0.0641, 0.3811)$ & 0.1574 & $(-0.0670, 0.3729)$ \\
		$\delta_{\text{SL}}$  & 0.3056 & $(0.0974, 0.5083)$  & 0.3044 & $(0.0966, 0.5053)$  & 0.2915 & $(0.0817, 0.4899)$ \\
		$\delta_{\text{ISO}}$ & 0.0693 & $(-0.0935, 0.2365)$ & 0.0688 & $(-0.0946, 0.2365)$ & 0.0679 & $(-0.0950, 0.2320)$ \\
		$R_{\text{AR}}$  & 1.4765 & $(0.7948, 2.2738)$ & 1.4756 & $(0.7868, 2.2627)$ & 1.4542 & $(0.7744, 2.2159)$ \\
		$R_{\text{SL}}$  & 1.8480 & $(1.0976, 2.7619)$ & 1.8425 & $(1.0754, 2.7339)$ & 1.8058 & $(1.0906, 2.7016)$ \\
		$R_{\text{ISO}}$ & 1.2255 & $(0.7651, 1.7973)$ & 1.2236 & $(0.7656, 1.7939)$ & 1.2192 & $(0.7601, 1.7774)$ \\
		$\psi_{\text{AR}}$  & 2.2293 & $(0.5386, 4.4521)$ & 2.2300 & $(0.5419, 4.4588)$ & 2.1540 & $(0.5362, 4.2701)$ \\
		$\psi_{\text{SL}}$  & 4.1562 & $(1.0520, 8.2511)$ & 4.1340 & $(1.0523, 8.1933)$ & 3.8705 & $(1.0013, 7.6229)$ \\
		$\psi_{\text{ISO}}$ & 1.4346 & $(0.5528, 2.5207)$ & 1.4314 & $(0.5747, 2.5285)$ & 1.4224 & $(0.5375, 2.4660)$ \\
		\bottomrule
	\end{tabular*}
	{\raggedright \small \textit{Notes:} DOM is the reference group. $\delta_i=\pi_i-\pi_1$, $R_i=\pi_i/\pi_1$, $\psi_i$ are relative to DOM for $i=$ AR, SL, ISO.\par}
\end{table}
\begin{table}[htbp]
	\centering
	\caption{Hypothesis test results for the RP data.}
	\label{tab:rp_tests}
	\begin{tabular*}{\textwidth}{@{\extracolsep\fill}lccc@{\extracolsep\fill}}
		\toprule
		\multirow{2}{*}{Quantity} & \multicolumn{3}{c}{Prior} \\
		\cmidrule(lr){2-4}
		& Reference & Jeffreys' & Uniform \\
		\midrule
		\multicolumn{4}{l}{\textit{(a) $H_{0i}:\delta_i=0$}}\\
		$BF_{01}^{\delta_{\text{AR}}}$  & 0.8678 & 0.8732 & 1.0064 \\
		$BF_{01}^{\delta_{\text{SL}}}$  & 0.0579 & 0.0621 & 0.0796 \\
		$BF_{01}^{\delta_{\text{ISO}}}$ & 2.2673 & 2.2903 & 2.4843 \\
		\multicolumn{4}{l}{\textit{(b) $H_0^*:\gamma_1=\gamma_2=\gamma_3=\gamma_4$}}\\
		Calibrated $\delta_0$ & 0.0740 & 0.0810 & 0.0660 \\
		$W$ posterior mean & 0.2637 & 0.2637 & 0.2622 \\
		$W$ 95\% HPD      & $(0.0697, 0.4777)$ & $(0.0697, 0.4777)$ & $(0.0676, 0.4748)$ \\
		\bottomrule
	\end{tabular*}
	{\raggedright \small \textit{Notes:} $BF_{01}$ is the Savage--Dickey Bayes factor for $H_{0i}:\delta_i=0$ vs.\ $H_{1i}:\delta_i\neq 0$. The posterior range $W=\max_i\gamma_i-\min_i\gamma_i$ is computed under the saturated model. The calibrated margins $\delta_0$ are averaged over $m$ for $g=4$ from Appendix~\ref{app:C}; since the RP data are unbalanced with $(m_1,\dots,m_4)=(28,21,19,148)$, no single calibrated value applies.\par}
\end{table}
\begin{table}[htbp]
	\centering
	\caption{Posterior means and 95\% HPD intervals of the group-specific correlation parameters $\gamma_i$ under the saturated model.}
	\label{tab:rp_saturated}
	\begin{tabular*}{\textwidth}{@{\extracolsep\fill}lcc@{\extracolsep\fill}}
		\toprule
		Group & Posterior mean & 95\% HPD \\
		\midrule
		DOM & 0.3122 & $(0.1052, 0.5323)$ \\
		AR  & 0.2315 & $(0.0667, 0.4164)$ \\
		SL  & 0.0815 & $(0.0046, 0.1813)$ \\
		ISO & 0.1766 & $(0.1104, 0.2461)$ \\
		\bottomrule
	\end{tabular*}
	{\raggedright \small \textit{Notes:} The posterior mean of $W=\max_i\gamma_i-\min_i\gamma_i$ is 0.2637 with 95\% HPD $(0.0697, 0.4777)$.\par}
\end{table}

Table~\ref{tab:rp_estimates} reports the posterior means and 95\% HPD
intervals under the three priors, which yield nearly identical estimates.
The marginal cure rates follow the ordering SL (0.6885--0.7013) \(>\) AR
(0.5545--0.5616) \(>\) ISO (0.4649--0.4657) \(>\) DOM (0.3957--0.3970). The common correlation parameter is estimated at \(\gamma\approx 0.1770\) with 95\% HPD \((0.1236, 0.2343)\), indicating moderate positive dependence between paired eyes. Relative to DOM, only SL shows a significant difference: the 95\% HPD intervals for \(\delta_{\text{SL}}\) \((0.0974, 0.5083)\), \(R_{\text{SL}}\) \((1.0976, 2.7619)\), and \(\psi_{\text{SL}}\) \((1.0520, 8.2511)\) all exclude their null values.
whereas those for AR and ISO contain zero or one. Table~\ref{tab:rp_tests} reports the hypothesis test results. For the individual risk differences, the Savage--Dickey Bayes factors under the reference prior are \(BF_{01}^{\delta_{\text{AR}}}=0.8678\),
\(BF_{01}^{\delta_{\text{SL}}}=0.0579\), and
\(BF_{01}^{\delta_{\text{ISO}}}=2.2673\); only SL provides strong evidence
against \(H_{0i}:\delta_i=0\), consistent with the HPD intervals above. 

To assess whether the four genetic groups share the same correlation structure, we fit the saturated model with group-specific \(\gamma_i\). Table~\ref{tab:rp_saturated} reports the posterior means and 95\% HPD intervals of \(\gamma_i\). The group-specific estimates are \((0.3122, 0.2315, 0.0815, 0.1766)\) for (DOM, AR, SL, ISO), corresponding to \((0.6878, 0.7685, 0.9185, 0.8234)\) under Li et al.'s parameterization, in close agreement with their MLEs. The posterior range \(W\), which measures the heterogeneity of correlations across groups, has a posterior mean of \(0.2637\) with 95\% HPD \((0.0697, 0.4777)\), excluding zero. Applying the HDI+ROPE rule with the calibrated margins for \(g=4\), the conclusion depends on the prior: under the uniform prior, \(L_W=0.0700>\delta_0=0.0660\), so \(H_0^*\) is rejected; under the Jeffreys' and Bernardo's reference priors, \(L_W\) falls below the corresponding margins (0.0810 and 0.0740) and \(H_0^*\) is not rejected. This sensitivity contrasts with the frequentist analysis of Li et al.~(2020)\cite{li2020statistical}, which did not reject equal correlations (\(T_{SC}=4.18\), \(p=0.24\)).

\section{Conclusions}\label{Sec7}
This paper proposed objective Bayesian methods for multi-group bilateral data under Dallal's model, covering both interval estimation and hypothesis testing. We derived three objective priors, namely the uniform, Jeffreys', and Bernardo's reference priors, and developed a posterior range test based on the HDI+ROPE decision rule for testing the homogeneity of correlations across groups. We calibrated the equivalence margin separately for each combination of group number, sample size, and prior.

Numerical studies first evaluated interval estimation using ETCP, EWCI, and EMSE. The Bayesian HPD intervals achieved coverage close to the nominal level and produced narrower intervals and smaller MSE than the Wald intervals, especially in small samples. The reference and Jeffreys' priors performed similarly, while the uniform prior was slightly more conservative. We then assessed the posterior range test in terms of empirical TIE and power. The Bayesian methods maintained empirical TIE close to the nominal \(5\%\) level. Bernardo's reference prior showed the most stable TIE control. The Jeffreys' prior was conservative, and the uniform prior was occasionally liberal. In contrast, the Wald test exhibited clear TIE inflation in small samples, but all methods performed similarly for large samples. In terms of power, the empirical values increased with sample size and with the correlation parameter. For small samples, the Wald test appeared to have higher power, but this was largely driven by its inflated TIEs. The Bernardo's reference prior maintained competitive power while keeping the TIE close to the nominal level. For large samples, all methods performed similarly, with powers approaching \(100\%\). Overall, the Bernardo's reference prior achieved the best balance between TIE control and power. The simulation study focuses on \(g=3\), while the appendix provides calibrated margins for \(g=4\) and \(g=5\) for future use.

We analyzed two real datasets to illustrate the proposed methods. The scleroderma trial with \(g=2\) confirmed that the proposed framework reduces to the original Dallal's model. The retinitis pigmentosa study with \(g=4\) showed the method's applicability to unbalanced multi-group designs. These examples also highlighted that the correlation homogeneity test can be sensitive to the choice of prior and calibrated margins.

Several limitations should be noted. Calibrating the equivalence margin is computationally intensive for larger group numbers. The posterior range test assumes a common correlation parameter under the null hypothesis. The simulation study focuses on balanced designs. Future research could extend the method to \(g\ge 4\) with a full simulation study, standardize the range statistic to reduce the dependence on group number, and incorporate covariates via Dallal's regression model.

\bmsection*{Author contributions}
Jinxiu Wen: Writing original draft, Methodology, Conceptualization. Zhiming Li: Writing review \& editing, Supervision, and Funding acquisition.

\bmsection*{Conflict of interest}
The authors declare no potential conflict of interest.

\bibliography{wileyNJD-AMA}

\appendix
\bmsection{Derivation of Jeffreys' Prior}
\label{app:A}
The first and
second derivatives of the log-likelihood function with respect to \(\gamma\) and \(u_i\) are
\begin{eqnarray*}\label{app:A3}
	\begin{aligned}
		\frac{\partial \ell}{\partial \gamma}&=\frac{\partial \ell(\gamma)}{\partial \gamma}=\frac{M_{1}}{\gamma}-\frac{M_{2}}{(1-\gamma)}-\frac{M_{1}+M_{2}}{(1+\gamma)}, \frac{\partial^2 \ell}{\partial \gamma^2}=-\frac{M_{1}}{\gamma^2}-\frac{M_{2}}{(1-\gamma)^2}+\frac{M_{1}+M_{2}}{(1+\gamma)^2},\\
		\frac{\partial \ell }{\partial u_i}&=\frac{\partial \ell(u_i)}{\partial u_i}=\frac{m_{1i}+m_{2i}}{u_i}-\frac{m_{0i}}{1-u_i}, \frac{\partial^2 \ell }{\partial u_i^2}=-\frac{m_{1i}+m_{2i}}{u_i^2}-\frac{m_{0i}}{(1-u_i)^2},\\
		\frac{\partial^2 \ell}{\partial\gamma\partial u_i}&= 0, \frac{\partial^2 \ell}{\partial u_i\partial u_{i'}}=0,\;i\ne i'.
	\end{aligned}
\end{eqnarray*}
Then, the Fisher information matrix \(I\) with respect to \(\gamma, \boldsymbol{u}\) is
\begin{eqnarray*}
	\begin{aligned}
		I(\gamma,\boldsymbol{u})=\text{diag}\left(I_{\gamma}, I_{u_1},\cdots,I_{u_i} \right),
	\end{aligned}
\end{eqnarray*}
where
\begin{eqnarray*}
	\begin{aligned}
		I_{\gamma}=E\Big[-\frac{\partial^2 \ell }{\partial \gamma^2}\Big]=\frac{2\sum_{i=1}^{g}m_{i} u_i}{\gamma(1-\gamma)(1+\gamma)^2}, I_{u_i}=E\Big[-\frac{\partial^2 \ell }{\partial u_i^2}\Big]=\frac{m_i}{u_i(1-u_i)}, \quad i=1,\cdots,g.
	\end{aligned}
\end{eqnarray*}
Define \(r_i=\frac{m_i}{m_1}\) (\(i=1, \cdots,g\)) be the ratio of the sample size in two groups, then \(m_{i}=m_{1}r_{i}\) and \(r_1=1\). Hence, the Jeffreys' prior under the parameterization \((\gamma,\boldsymbol{u})\) is
\begin{eqnarray*}
	\begin{aligned}
		\pi_J(\gamma,\boldsymbol{u})
		\propto \sqrt{\det I(\gamma,\boldsymbol{u})}
		\propto\sqrt{\frac{ \sum_{i=1}^{g}r_{i} u_i}{\gamma(1-\gamma)(1+\gamma)^2 \prod_{i=1}^g u_i(1-u_i)}}, \quad 0<\gamma, u_i<1.
	\end{aligned}
\end{eqnarray*}
Then, the Jeffreys' prior density for \((\gamma,\boldsymbol{\pi})\) is
\begin{eqnarray*}
	\begin{aligned}
		\pi_J(\gamma,\boldsymbol{\pi})
		\propto \sqrt{\frac{\sum_{i=1}^{g}r_{i} \pi_i}{\gamma(1-\gamma)(1+\gamma)^{1-g}\prod_{i=1}^g \pi_i[1-(1+\gamma)\pi_i]}}, \quad (\gamma, \boldsymbol{\pi}) \in \Omega.
	\end{aligned}
\end{eqnarray*}

\bmsection{Calibrated Equivalence Margins for \(g=4\) and \(g=5\)}
\label{app:B}
\begin{table}[htbp]
	\centering
	\caption{Calibrated equivalence margins \(\delta_0\) for the posterior range test with \(g=4\) and \(g=5\). Each value is obtained by simulating under \(H_0:\delta_2=\cdots=\delta_g\) and selecting the smallest \(\delta_0\) such that the empirical TIE rate equals \(0.05\).}
	\label{tab:calibration_appendix}
	\begin{tabular*}{\textwidth}{@{\extracolsep\fill}lccccccccc@{\extracolsep\fill}}
		\toprule
		\multirow{2}{*}{\(g\)} & \multirow{2}{*}{Prior} & \multicolumn{4}{c}{\(\delta_0\)} & \multicolumn{4}{c}{Empirical TIE} \\
		\cmidrule(lr){3-6} \cmidrule(lr){7-10}
		& & \(m=10\) & \(m=25\) & \(m=50\) & \(m=100\) & \(m=10\) & \(m=25\) & \(m=50\) & \(m=100\) \\
		\midrule
		\multirow{3}{*}{4} & Uniform   & 0.1034 & 0.0619 & 0.0554 & 0.0430 & 0.050 & 0.050 & 0.050 & 0.052 \\
		& Jeffreys  & 0.1379 & 0.0779 & 0.0649 & 0.0420 & 0.050 & 0.050 & 0.050 & 0.050 \\
		& Reference & 0.1164 & 0.0839 & 0.0599 & 0.0370 & 0.050 & 0.050 & 0.052 & 0.050 \\
		\midrule
		\multirow{3}{*}{5} & Uniform   & 0.1738 & 0.1104 & 0.0814 & 0.0584 & 0.050 & 0.050 & 0.050 & 0.050 \\
		& Jeffreys  & 0.2283 & 0.1194 & 0.0849 & 0.0564 & 0.050 & 0.050 & 0.050 & 0.052 \\
		& Reference & 0.2068 & 0.1319 & 0.0869 & 0.0619 & 0.050 & 0.050 & 0.048 & 0.050 \\
		\bottomrule
	\end{tabular*}
	{\raggedright \small \textit{Notes:} Each value of \(\delta_0\) is calibrated via \(N=500\) simulated datasets under the null hypothesis. The empirical TIE is computed as the proportion of datasets in which the 95\% HPD lower bound of the posterior range \(W\) exceeds \(\delta_0\).\par}
\end{table}

\bmsection{Additional Simulation Results for interval estimation}
\label{app:C}
\begin{sidewaystable}[htbp]
	\centering
	\caption{Performance of 95\% HPDs/CIs for the common risk difference for $\pi_{1}=0.3$ with \(g=3\).}
	\label{tab:ci_pi0.3}
	\begin{tabular*}{\textwidth}{@{\extracolsep\fill}lccccccccccccccccccc@{\extracolsep\fill}}
		\toprule
		\multirow{2}{*}{} & \multirow{2}{*}{$\delta$} & \multirow{2}{*}{$\pi_1$} & \multirow{2}{*}{$\gamma$} &  \multicolumn{4}{c}{$\boldsymbol{m}=(10,10,10)$} & \multicolumn{4}{c}{$\boldsymbol{m}=(25,25,25)$} & \multicolumn{4}{c}{$\boldsymbol{m}=(50,50,50)$} & \multicolumn{4}{c}{$\boldsymbol{m}=(100,100,100)$} \\
		\cmidrule(lr){5-8} \cmidrule(lr){9-12} \cmidrule(lr){13-16} \cmidrule(lr){17-20}
		& & & & $\text{CI}_\text{U}$ & $\text{CI}_\text{J}$ & $\text{CI}_\text{R}$ & $\text{CI}_\text{W}$ & $\text{CI}_\text{U}$ & $\text{CI}_\text{J}$ & $\text{CI}_\text{R}$ & $\text{CI}_\text{W}$ & $\text{CI}_\text{U}$ & $\text{CI}_\text{J}$ & $\text{CI}_\text{R}$ & $\text{CI}_\text{W}$ & $\text{CI}_\text{U}$ & $\text{CI}_\text{J}$ & $\text{CI}_\text{R}$ & $\text{CI}_\text{W}$ \\
		\midrule
		\multirow{9}{*}{METCP} 
		& \multirow{3}{*}{0.0}
		& \multirow{3}{*}{II} & $a_1$ 
		& 96.00 & 93.40 & 93.50 & 90.70 & 95.40 & 93.80 & 93.80 & 91.00 & 95.20 & 93.80 & 93.80 & 90.90 & 95.20 & 95.00 & 95.00 & 94.80 \\ 
		& & & $a_2$ 
		& 94.80 & 94.40 & 94.30 & 94.30 & 95.50 & 94.40 & 94.20 & 94.10 & 95.30 & 95.20 & 95.20 & 94.90 & 94.70 & 94.40 & 94.50 & 94.30 \\ 
		& & & $a_3$ 
		& 94.70 & 94.50 & 94.60 & 94.50 & 94.70 & 94.40 & 94.50 & 94.30 & 95.90 & 95.80 & 95.80 & 95.80 & 94.60 & 94.50 & 94.50 & 94.40 \\  
		& \multirow{3}{*}{0.1}
		& \multirow{3}{*}{II} & $a_1$ 
		& 95.10 & 94.20 & 94.00 & 91.10 & 95.40 & 94.20 & 94.40 & 92.20 & 94.80 & 93.10 & 93.10 & 92.50 & 95.20 & 94.60 & 94.60 & 94.00 \\ 
		& & & $a_2$ 
		& 95.50 & 94.90 & 94.60 & 93.90 & 96.30 & 96.10 & 96.00 & 95.60 & 95.30 & 95.00 & 94.90 & 94.80 & 96.00 & 96.10 & 96.10 & 95.80 \\ 
		& & & $a_3$ 
		& 95.30 & 95.20 & 95.00 & 95.00 & 94.90 & 94.80 & 94.90 & 94.80 & 95.10 & 95.00 & 94.90 & 94.90 & 95.20 & 94.80 & 95.20 & 94.80 \\  
		& \multirow{3}{*}{0.3}
		& \multirow{3}{*}{II} & $a_1$ 
		& 95.00 & 94.40 & 94.40 & 91.70 & 94.50 & 93.70 & 93.70 & 91.80 & 93.20 & 92.80 & 93.10 & 90.50 & 94.70 & 94.20 & 94.20 & 93.90 \\ 
		& & & $a_2$ 
		& 94.00 & 94.00 & 94.00 & 93.80 & 94.70 & 94.10 & 94.30 & 94.20 & 94.80 & 94.90 & 94.90 & 94.80 & 95.50 & 95.40 & 95.40 & 94.70 \\ 
		& & & $a_3$ 
		& 94.80 & 94.50 & 94.60 & 93.80 & 94.90 & 94.90 & 94.90 & 94.50 & 94.70 & 94.60 & 94.60 & 94.30 & 95.00 & 95.20 & 95.20 & 94.60 \\  
		\multirow{9}{*}{ETCP} 
		& \multirow{3}{*}{0.0}
		& \multirow{3}{*}{II} & $a_1$ 
		& 93.00 & 88.50 & 88.80 & 84.20 & 91.70 & 89.10 & 88.90 & 84.10 & 91.30 & 89.00 & 89.20 & 84.30 & 91.20 & 90.70 & 90.90 & 90.60 \\ 
		& & & $a_2$ 
		& 90.50 & 89.90 & 89.80 & 89.80 & 92.10 & 90.30 & 90.00 & 89.80 & 91.40 & 91.20 & 91.30 & 90.80 & 90.10 & 89.60 & 89.70 & 89.50 \\ 
		& & & $a_3$ 
		& 90.50 & 90.40 & 90.40 & 90.30 & 90.20 & 89.80 & 90.00 & 89.80 & 92.40 & 92.40 & 92.30 & 92.40 & 90.40 & 90.20 & 90.20 & 90.10 \\ 
		& \multirow{3}{*}{0.1}
		& \multirow{3}{*}{II} & $a_1$ 
		& 90.90 & 89.80 & 89.40 & 84.30 & 91.80 & 89.90 & 90.30 & 86.90 & 90.60 & 87.80 & 87.50 & 86.80 & 91.70 & 90.70 & 90.90 & 89.90 \\ 
		& & & $a_2$ 
		& 92.00 & 90.90 & 90.20 & 88.90 & 93.20 & 92.90 & 92.70 & 91.90 & 91.60 & 91.00 & 90.80 & 90.60 & 92.80 & 92.80 & 92.90 & 92.30 \\ 
		& & & $a_3$ 
		& 91.70 & 91.20 & 91.10 & 90.90 & 90.70 & 90.60 & 90.80 & 90.50 & 90.90 & 90.90 & 90.70 & 90.70 & 91.20 & 90.40 & 91.20 & 90.30 \\  
		& \multirow{3}{*}{0.3}
		& \multirow{3}{*}{II} & $a_1$ 
		& 91.30 & 90.40 & 90.30 & 85.80 & 91.30 & 89.60 & 89.60 & 86.40 & 90.00 & 89.20 & 89.30 & 86.90 & 90.60 & 89.70 & 89.90 & 89.10 \\ 
		& & & $a_2$ 
		& 89.90 & 90.00 & 89.90 & 89.30 & 91.30 & 89.90 & 90.10 & 90.10 & 90.40 & 90.70 & 90.60 & 90.60 & 92.30 & 92.00 & 92.10 & 91.10 \\ 
		& & & $a_3$ 
		& 91.80 & 90.90 & 91.30 & 89.40 & 90.90 & 90.80 & 90.70 & 89.90 & 91.00 & 90.70 & 90.70 & 90.20 & 91.40 & 91.40 & 91.30 & 90.60 \\  
		\multirow{6}{*}{EWCI} 
		& \multirow{2}{*}{0.0}
		& \multirow{2}{*}{II} & $a_1$ 
		& 0.596 & 0.618 & 0.613 & 0.664 & 0.565 & 0.582 & 0.579 & 0.625 & 0.505 & 0.516 & 0.514 & 0.552 & 0.412 & 0.420 & 0.418 & 0.434 \\ 
		& & & $a_2$ 
		& 0.389 & 0.395 & 0.394 & 0.408 & 0.345 & 0.349 & 0.348 & 0.361 & 0.302 & 0.306 & 0.305 & 0.311 & 0.284 & 0.286 & 0.285 & 0.291 \\ 
		& & & $a_3$ 
		& 0.252 & 0.253 & 0.253 & 0.258 & 0.217 & 0.218 & 0.218 & 0.220 & 0.204 & 0.205 & 0.205 & 0.207 & 0.181 & 0.181 & 0.181 & 0.183 \\
		& \multirow{2}{*}{0.1}
		& \multirow{2}{*}{II} & $a_1$ 
		& 0.605 & 0.627 & 0.625 & 0.681 & 0.568 & 0.586 & 0.584 & 0.634 & 0.500 & 0.509 & 0.510 & 0.548 & 0.421 & 0.429 & 0.428 & 0.446 \\ 
		& & & $a_2$ 
		& 0.392 & 0.398 & 0.397 & 0.412 & 0.344 & 0.347 & 0.348 & 0.360 & 0.308 & 0.311 & 0.311 & 0.318 & 0.287 & 0.290 & 0.289 & 0.295 \\ 
		& & & $a_3$ 
		& 0.250 & 0.251 & 0.251 & 0.256 & 0.222 & 0.223 & 0.223 & 0.225 & 0.207 & 0.207 & 0.207 & 0.210 & 0.180 & 0.180 & 0.180 & 0.182 \\ 
		& \multirow{2}{*}{0.3}
		& \multirow{2}{*}{II} & $a_1$ 
		& 0.586 & 0.600 & 0.600 & 0.649 & 0.540 & 0.548 & 0.549 & 0.587 & 0.446 & 0.442 & 0.443 & 0.462 & 0.404 & 0.408 & 0.409 & 0.424 \\ 
		& & & $a_2$ 
		& 0.368 & 0.370 & 0.371 & 0.383 & 0.298 & 0.297 & 0.297 & 0.303 & 0.294 & 0.296 & 0.296 & 0.302 & 0.266 & 0.267 & 0.268 & 0.272 \\ 
		& & & $a_3$ 
		& 0.214 & 0.213 & 0.213 & 0.216 & 0.212 & 0.213 & 0.213 & 0.215 & 0.192 & 0.192 & 0.192 & 0.194 & 0.153 & 0.153 & 0.153 & 0.154 \\  
		\multirow{6}{*}{EMSE} 
		& \multirow{2}{*}{0.0}
		& \multirow{2}{*}{II} & $a_1$ 
		& 0.0221 & 0.0270 & 0.0264 & 0.0325 & 0.0198 & 0.0237 & 0.0234 & 0.0284 & 0.0165 & 0.0193 & 0.0191 & 0.0227 & 0.0104 & 0.0113 & 0.0112 & 0.0122 \\ 
		& & & $a_2$ 
		& 0.0093 & 0.0101 & 0.0100 & 0.0108 & 0.0073 & 0.0078 & 0.0077 & 0.0083 & 0.0058 & 0.0061 & 0.0060 & 0.0063 & 0.0051 & 0.0053 & 0.0053 & 0.0055 \\ 
		& & & $a_3$ 
		& 0.0042 & 0.0043 & 0.0043 & 0.0044 & 0.0032 & 0.0032 & 0.0032 & 0.0033 & 0.0026 & 0.0027 & 0.0027 & 0.0027 & 0.0021 & 0.0022 & 0.0021 & 0.0022 \\  
		& \multirow{2}{*}{0.1}
		& \multirow{2}{*}{II} & $a_1$ 
		& 0.0241 & 0.0285 & 0.0283 & 0.0344 & 0.0201 & 0.0238 & 0.0236 & 0.0288 & 0.0167 & 0.0191 & 0.0191 & 0.0226 & 0.0113 & 0.0122 & 0.0122 & 0.0132 \\ 
		& & & $a_2$ 
		& 0.0099 & 0.0107 & 0.0106 & 0.0115 & 0.0070 & 0.0074 & 0.0074 & 0.0080 & 0.0060 & 0.0063 & 0.0062 & 0.0065 & 0.0050 & 0.0052 & 0.0052 & 0.0054 \\ 
		& & & $a_3$ 
		& 0.0039 & 0.0040 & 0.0040 & 0.0042 & 0.0032 & 0.0033 & 0.0033 & 0.0034 & 0.0027 & 0.0027 & 0.0027 & 0.0028 & 0.0021 & 0.0021 & 0.0021 & 0.0022 \\ 
		& \multirow{2}{*}{0.3}
		& \multirow{2}{*}{II} & $a_1$ 
		& 0.0237 & 0.0258 & 0.0258 & 0.0302 & 0.0214 & 0.0229 & 0.0229 & 0.0264 & 0.0145 & 0.0152 & 0.0152 & 0.0172 & 0.0110 & 0.0114 & 0.0114 & 0.0121 \\ 
		& & & $a_2$ 
		& 0.0093 & 0.0096 & 0.0096 & 0.0101 & 0.0059 & 0.0059 & 0.0059 & 0.0062 & 0.0058 & 0.0059 & 0.0059 & 0.0061 & 0.0045 & 0.0046 & 0.0046 & 0.0048 \\ 
		& & & $a_3$ 
		& 0.0031 & 0.0031 & 0.0031 & 0.0032 & 0.0030 & 0.0031 & 0.0031 & 0.0031 & 0.0024 & 0.0025 & 0.0025 & 0.0025 & 0.0015 & 0.0015 & 0.0015 & 0.0015 \\   
		\bottomrule
	\end{tabular*}%
\end{sidewaystable}

\end{document}